\documentclass[11pt,a4paper]{article}
 \pdfoutput=1
    \usepackage{jheppub}

\usepackage{ifpdf}

\usepackage{afterpage}
\usepackage[utf8]{inputenc}

\usepackage{amssymb}
\usepackage{amsmath}
\usepackage{graphicx}
\usepackage{longtable}
\usepackage{verbatim}
\usepackage{amsfonts}
\usepackage{xcolor}

\DeclareMathOperator{\sign}{sign}
\makeatletter
\newcommand{\vast}{\bBigg@{4}}
\newcommand{\Vast}{\bBigg@{5}}
\makeatother

\newcommand{\bear}{\begin{array}}
\newcommand{\ear}{\end{array}}

\newcommand{\beq}{\begin{eqnarray}}
\newcommand{\eeq}{\end{eqnarray}}
\newcommand{\beqa}{\begin{eqnarray}}
\newcommand{\eeqa}{\end{eqnarray}}

\def\OMIT#1{{}}
\newcommand{\lsim}{\mathrel{\rlap{\lower4pt\hbox{\hskip1pt$\sim$}}
    \raise1pt\hbox{$<$}}}         
\newcommand{\gsim}{\mathrel{\rlap{\lower4pt\hbox{\hskip1pt$\sim$}}
    \raise1pt\hbox{$>$}}}         

\newcommand{\be}{\begin{equation}}
\newcommand{\ee}{\end{equation}}
\newcommand{\ba}{\begin{eqnarray}}
\newcommand{\ea}{\end{eqnarray}}

\def\lsim{\mathrel{\rlap{\lower4pt\hbox{\hskip1pt$\sim$}}
    \raise1pt\hbox{$<$}}}         
\def\gsim{\mathrel{\rlap{\lower4pt\hbox{\hskip1pt$\sim$}}
    \raise1pt\hbox{$>$}}}         

\newcommand{\Ai}{\operatorname{Ai}}
\newcommand{\Bi}{\operatorname{Bi}}

\renewcommand{\tilde}{\widetilde}

\vspace*{-30mm}

\title{\boldmath An Inflationary Probe of Cosmic Higgs Switching}
\author[a]{JiJi Fan,} 
\author[b]{Matthew Reece,}
\author[c,d]{and Yi Wang}
\affiliation[a]{Department of Physics, Brown University, Providence, RI, 02912, USA}
\affiliation[b]{Department of Physics, Harvard University, Cambridge, MA 02138, USA}
\affiliation[c]{Department of Physics, The Hong Kong University of Science and Technology, Clear Water Bay, Kowloon, Hong Kong, P.R.China}
\affiliation[d]{Jockey Club Institute for Advanced Study, The Hong Kong University of Science and Technology, Clear Water Bay, Kowloon, Hong Kong, P.R.China}

\vspace*{1cm}

\abstract{
A scalar Higgs field can be repeatedly switched on and off when it couples to a classically oscillating scalar modulus field. The modulus flips the Higgs mass term between stable and tachyonic values. We study a cosmological scenario in which such repeated phase transitions occur during inflation. An irrelevant operator coupling the Higgs field to the inflaton can then imprint the pattern of phase transitions in the correlation functions of the inflaton. Using both numerical and analytic studies, we show that the inflaton 2-point function carries characteristic imprints of the modulus oscillation and its effect on the Higgs boson. We briefly remark on the potential observability of such patterns and how they might be distinguished from other dynamics in the early universe. 
    
}
\begin{document}
\maketitle

\section{Introduction}

The discovery of the Higgs boson is a milestone in particle physics and marks the completion of the standard model (SM). Yet the origin of the Higgs potential and its related early Universe dynamics remains mysterious and are among the deepest puzzles in fundamental physics. In particular, there is an intriguing possibility that the parameters of the Higgs potential (in general, the SM parameters) are dynamical in the early Universe, resulting in interesting phenomena before settling down to fixed values we could measure today. In fact, such a possibility arises naturally when we invoke new mechanism beyond the SM to explain the origin of the Higgs mass, either in a natural or fine-tuned way. For instance, in supersymmetry (SUSY),\footnote{Experimental results from the Large Hadron Collider (LHC) suggest that if SUSY is realized in nature, it is most likely that its associated electroweak symmetry breaking is tuned at a few percent level or worse.} SM parameters are not truly constants but are controlled by values of some other scalar fields which are usually referred to as moduli. A modulus field could oscillate in the early Universe and lead to a varying Higgs mass parameter through its coupling to the Higgs field. More specifically, given a trilinear coupling between a modulus $\phi$ and the Higgs field $h$, e.g., $\phi h^\dagger h$, the Higgs mass could flip sign when the modulus oscillates. On the positive side of the modulus, the Higgs mass parameter is positive and electroweak symmetry is unbroken while on the negative side, the Higgs mass is negative and electroweak symmetry is broken. Thus as the modulus oscillates, the Higgs field could oscillate between two different phases. It has been pointed out before that if certain parametric relations are satisfied, this feature could lead to a new epoch in the early Universe featuring violent dynamics with Higgs particle production and rapid modulus fragmentation, resulting in primordial gravitational wave production~\cite{Amin:2019qrx}.\footnote{Another example of a time-dependent Higgs potential, aiming at providing a technically natural explanation of the Higgs mass, is the relaxion scenario~\cite{Graham:2015cka}. In the relaxion mechanism, the modulus field scans the Higgs mass while rolling down its potential. It doesn't oscillate.}

In this article, we will explore a different possible cosmological signal from the phase transition oscillations of the Higgs field in the early Universe. If the Higgs couples to the inflaton, its oscillations between unbroken and broken symmetry phases will imprint on the primordial inflaton spectrum. 

In the literature, an oscillating field with a constant mass in the primordial epoch is referred to as a ``Standard Clock". It oscillates with a frequency that could be thought of as ``ticks" of a clock~\cite{Chen:2011zf}. The oscillations could resonate with the inflaton background and imprint the ticks as special types of oscillating patterns in the primordial inflaton spectrum \cite{Chen:2011zf,Shiu:2011qw,Saito:2012pd,Gao:2013ota,Noumi:2013cfa}.\footnote{Earlier studies of resonance phenomena in density perturbations, outside the context of Standard Clocks, include \cite{Chen:2008wn,Flauger:2009ab,Flauger:2010ja,Chen:2010bka}. An earlier study of how a transient oscillation of a massive field can affect inflationary correlation functions was \cite{Burgess:2002ub}, which studied oscillations at the beginning rather than the middle of inflation and did not give the resonance interpretation. These oscillations are on scales too large to search for in the CMB, and give only a low-frequency modulation.} These patterns encode the time dependence of the scale factor, which could be used to distinguish inflation from alternative scenarios for the origin of the Universe~\cite{Chen:2011zf, Chen:2012ja, Chen:2014joa, Chen:2014cwa, Huang:2016quc}.\footnote{Even when classical oscillations of a massive field are absent, its quantum fluctuations could still lead to interesting oscillating features in the bi-spectrum. Thus measurements of non-Gausianity could be used to probe masses and spins of heavy particles during inflation (``quasi-single field inflation"~\cite{Chen:2009we, Chen:2009zp,Baumann:2011nk} and ``cosmological collider" physics~\cite{Arkani-Hamed:2015bza}) and distinguish inflation and alternatives (``quantum primordial clock"~\cite{Chen:2015lza}).}

In the scenario we study, with Higgs oscillation between unbroken and broken symmetry phases, the Higgs mass has a characteristic varying pattern. We will show that the oscillations between different phases (as opposed to no phase transition in the oscillations of the heavy field studied in the literature) give rise to non-trivial new ``k-wavepacket" features in the primordial spectrum. The oscillation pattern in phase and amplitude could potentially allow us to probe new energy scales above the weak scale and Higgs dynamics in the early Universe. It may even give us some hints about fine-tuning in the Higgs sector. 

The paper is organized as follows. In Sec.~\ref{sec:model}, we present the main simplified model we focus on, which contains an oscillating modulus, a Higgs field and an inflaton field. We describe the time evolution of the modulus and the Higgs field. In Sec.~\ref{sec:primordialspectrum}, we compute the corrections to the primordial perturbation spectrum due to the oscillations of the Higgs field between different phases. When the coupling between the inflaton and the Higgs could be treated as a perturbation, we derive some analytical understanding of the correction to the inflaton spectrum and compare them with the numerical results. In Sec.~\ref{sec:CMB}, we numerically compute the resulting CMB temperature anisotropy spectrum from the modified primordial spectrum. In Sec.~\ref{sec:compare}, we study the uniqueness of the signal and compare predictions from the phase oscillating model with a model in which the Higgs is always in the broken-symmetry phase. We conclude in Sec.~\ref{sec:con}. 

\section{The toy model}
\label{sec:model}

In this section, we present the model with the phase transition oscillations of the Higgs field. The setup contains the inflaton $\phi$, the Higgs $h$ (we only consider the radial mode\footnote{Only the radial mode matters for our following analysis since only $|h|$ couples to the inflaton.}) and a modulus field $\chi$. The inflationary background is approximately de Sitter with the scale factor $a(t)=e^{Ht}$. The metric convention is $(-, +, +, +)$. The matter Lagrangian is
\begin{align}
    \mathcal{L} = \mathcal{L}_{\phi} + \mathcal{L}_{h} + \mathcal{L}_{\chi}
                + \mathcal{L}_{\phi h} + \mathcal{L}_{\chi h}~,
     \label{eq:Lagrangian}
\end{align}
where 
\begin{align}
    \mathcal{L}_\phi = \sqrt{-g}
    \left  [ 
        - \frac{1}{2} (\partial\phi)^2 - V(\phi)
    \right ]~,
\end{align}
\begin{align}
    \mathcal{L}_h = \sqrt{-g} 
    \left  [ 
        - \frac{1}{2} (\partial h)^2 + \frac{1}{2}  m_h^2 h^2 - \frac{\lambda}{4} h^4
    \right ]~,
\end{align}
\begin{align}
    \mathcal{L}_\chi = \sqrt{-g}
    \left  [
        - \frac{1}{2} (\partial\chi)^2 - \frac{1}{2} m_\chi^2 \chi^2
    \right ]~,
\end{align}
\begin{align}
    \mathcal{L}_{\phi h} = \sqrt{-g} \left[-\frac{1}{2} \frac{y}{\Lambda^2} h^2 (\partial\phi)^2\right]~,
    \label{eq:phih}
\end{align}
\begin{align}
    \mathcal{L}_{\chi h} = \sqrt{-g} \left[- \frac{M^2}{2 f} \chi h^2 \right]~,
    \label{eq:chih}
\end{align}
where $g$ is the metric. In the model, there are a few energy scales: $m_\chi$ is the modulus mass; $f$ is the field range of the modulus; $m_h^2$ is the standard model Higgs mass squared parameter and $M^2$ sets the natural Higgs mass scale when $|\chi| \approx f$. In the tuned case, which we focus on, $M^2\gg m_h^2$. Thus in the early Universe with the modulus field present, the $\chi h^2$ coupling determines the Higgs mass, which varies when the value of $\chi$ changes. The modulus-inflaton coupling $y$ is dimensionless and is taken to be of ${\cal O}(1)$. 

The equations of motion for inflaton $\phi$, Higgs $h$ and modulus $\chi$ are then
\begin{align}\label{eq:eom-phi}
    \ddot\phi + 3H\dot\phi + V' 
    +\underline {\frac{y}{\Lambda^2} a^{-3} \partial_t \left ( h^2\dot\phi a^3 \right ) }
    = 0~, 
\end{align}
\begin{align}\label{eq:eom-h}
    \ddot h + 3H \dot h +
    \left  [
        \left ( M^2 \frac{\chi}{f} - m_h^2  \right ) 
      \underline{ - \frac{y}{\Lambda^2}\dot\phi^2 }
    \right ] h + \lambda h^3 = 0 ~,
    \end{align}
\begin{align}\label{eq:eom-chi}
    \ddot\chi + 3H \dot\chi + m_\chi^2 \chi 
\underline{+ \frac{1}{2f}M^2 h^2 }
    =0 ~.
\end{align}
We assume the back-reaction between different fields is negligible and thus, to the leading order, the underlined terms in the equations above could be neglected in the homogeneous solution. (We do include the underlined term in \eqref{eq:eom-phi} when considering the effect of $h$ on the $\phi$ perturbations, below.) These include ignoring back-reaction from
    \begin{itemize}
        \item The Higgs on the inflaton $\phi$, i.e.~the underlined term in Eq.~\eqref{eq:eom-phi} is not important. Naively, this appears to require $\Lambda \gg \sqrt{M^3/H}$. However, note that the underlined term in Eq.~\eqref{eq:eom-phi} oscillates quickly as $h$ oscillates. Thus the actual impact on $\phi$ may be smoothed out and the condition could be relaxed a bit. This is indeed evident in the numerical calculation. Thus, we will here assume the numerical constraint that the correction to the inflaton spectrum is much smaller than $\mathcal{O}(1)$. 
        \item The inflaton $\phi$ on the Higgs, i.e.~the underlined term in Eq.~\eqref{eq:eom-h} is negligible. This requires $\Lambda \gg \dot\phi / m_h$.
        \item The Higgs on the modulus $\chi$, i.e.~the underlined term in Eq.~\eqref{eq:eom-chi} is negligible. This requires $f\gg M^2/m_\chi$. In fact, this constraint may be further relaxed if there is a separation of scales $M\gg m_\chi$ since $h^2$ oscillates. Nevertheless, for simplicity we will restrict our attention to the parameter region satisfying $f\gg M^2/m_\chi$.
    \end{itemize}

In addition, we assume that
\begin{itemize}
    \item The modulus starts rolling from the symmetry breaking side. Thus the Higgs field configuration is dominated by the zero mode. Otherwise, the Higgs evolution trajectory bifurcates between different parts of the Universe and the dynamics becomes more complicated, similar to the case of multi-stream inflation \cite{Li:2009sp, Li:2009me, Abolhasani:2011yp}.
    \item The energy density of $\chi$ is subdominant compared to that of the inflaton. This requires $f\ll \sqrt{3}M_{\rm pl}H/m_\chi$ with $M_{\rm pl}$ being the reduced Planck scale. When this is satisfied, the Higgs energy density is also subdominant, because $\rho_h\sim M^4 \ll m_\chi^2 f^2$.
\end{itemize}

We consider the following hierarchy 
\begin{align}
|\chi_0| \sim f \gg M \gg m_\chi \gg m_h \gtrsim H,
\end{align}
where $\chi_0$ is the initial amplitude of the modulus and $H$ is the Hubble scale of the inflation. 
For example, one benchmark model we keep using in the rest of the paper has $M=1020H$, $m_\chi=10H$, $m_h=2H$. Given the observed normalization of the scalar perturbation spectrum $P_\zeta=(H^2/(2\pi\dot\phi))^2\simeq 2.4\times 10^{-9}$, we have $\dot\phi\simeq 3200H^2$. To satisfy the assumptions above except for the first one $\Lambda \gg \sqrt{M^3/H}$, we need $\Lambda \gg 1.6\times 10^3 H$ and $10^5H \ll f \ll 0.17 M_{\rm pl}$. We numerically find that even when $\Lambda = 6 \times 10^3 H$, which doesn't satisfy $\Lambda \gg \sqrt{M^3/H}$, the correction to the inflaton spectrum is still a perturbation. 

For simplicity we take the Higgs self-coupling $\lambda\sim \mathcal{O}(1)$ when estimating parameters.\footnote{The Higgs self-coupling is about 0.16 today. Yet in the early Universe, it could also vary and depend on the modulus field value, as explored in Ref.~\cite{Amin:2019qrx}.} We also set the coupling between inflaton and Higgs $y=1$ (a general $y\sim \mathcal{O}(1)$ can be absorbed into a redefinition of $\Lambda$).

To solve the system of equations, we start from Eq.~\eqref{eq:eom-chi} and solve the modulus' motion first. Given the assumption that the Higgs has negligible back-reaction on $\chi$, we have, up to an unimportant phase,
\begin{align}
    \chi \approx \chi_0 a^{-3/2} \cos(m_\chi t)~.
\end{align}
To have the oscillations between different phases happen classically, the modulus $\chi$ needs to oscillate, preferably starting from an amplitude $\chi_0\sim - f$. The oscillation can happen if the potential of $\chi$ is initially flat before falling to the quadratic part \cite{Chen:2014cwa}.\footnote{There are many other possibilities to trigger the modulus' oscillations, for example, a sharp turn in the trajectory between the inflaton and modulus direction, see for example, Ref.~\cite{Gao:2012uq}; or the Hubble scale decreases and falls below $m_\chi$, which may not be naturally compatible with our assumptions such as decoupled $\chi$-$\phi$ sectors and a large modulus mass.}

In our model, we ignore the direct coupling between the inflaton and the modulus. In principle, even if the coupling is absent at tree level, it could be generated radiatively. More specifically, couplings in Eq.~\eqref{eq:phih} and~\eqref{eq:chih} could generate at one-loop level an operator $(\partial \phi)^2 \chi$ with a coefficient of order $\frac{y}{16\pi^2} \frac{M^2}{\Lambda^2 f}$ up to a logarithmic factor. Then the contribution from the modulus-inflaton coupling to the modification of the inflaton kinetic term $(\partial \phi)^2$ is of order $\frac{y}{16\pi^2} \frac{M^2}{\Lambda^2}$, which is one loop factor suppressed compared to the contribution from the coupling between the Higgs and inflaton. In addition, since the modulus is always in the zero mode, the correction to the primordial spectrum is very similar to the classical primordial clock, which takes the form $\sin(C \log k)$ (with $C$ some constant)~\cite{Chen:2011zf}. The corrections to the primordial spectrum due to different couplings are additive. We will focus on the one from the coupling between the Higgs and inflaton below since, in part of the parameter space, it could be larger and leads to interesting new results. Future studies could examine the feasibility of disentangling these effects from those of a direct inflaton--modulus coupling when both are present.

\subsection{The evolution of the Higgs}
\label{sec:higgsprofile}

The Higgs mode can be split into two components, $h=h_\mathrm{vev}+h_\mathrm{osc}$, where $h_\mathrm{vev}$ is the instant Higgs vacuum expectation value (vev) obtained by minimizing the instantaneous Higgs potential (including the contribution from interacting with the modulus $\chi$), and $h_\mathrm{osc}$ is the oscillation on top of that. The effective Higgs mass squared is
\begin{align}
    m_\mathrm{eff}^2(t) = M^2 \frac{\chi_0}{f} e^{-\frac{3Ht}{2} } \cos(m_\chi t) - m_h^2~.
\end{align}
Below, we will assume for simplicity that oscillations begin at $t = 0$ when the modulus field has the value $\chi_0 =-f$. This is consistent with our assumption that the modulus starts rolling from the symmetry breaking side and Higgs is initially in the zero mode. $|m_\mathrm{eff}^2(t)|$ is initially  of order $M^2$ most of the time, then it gradually redshifts.

\begin{figure}[htbp] \centering
    \includegraphics[width=0.45\textwidth]{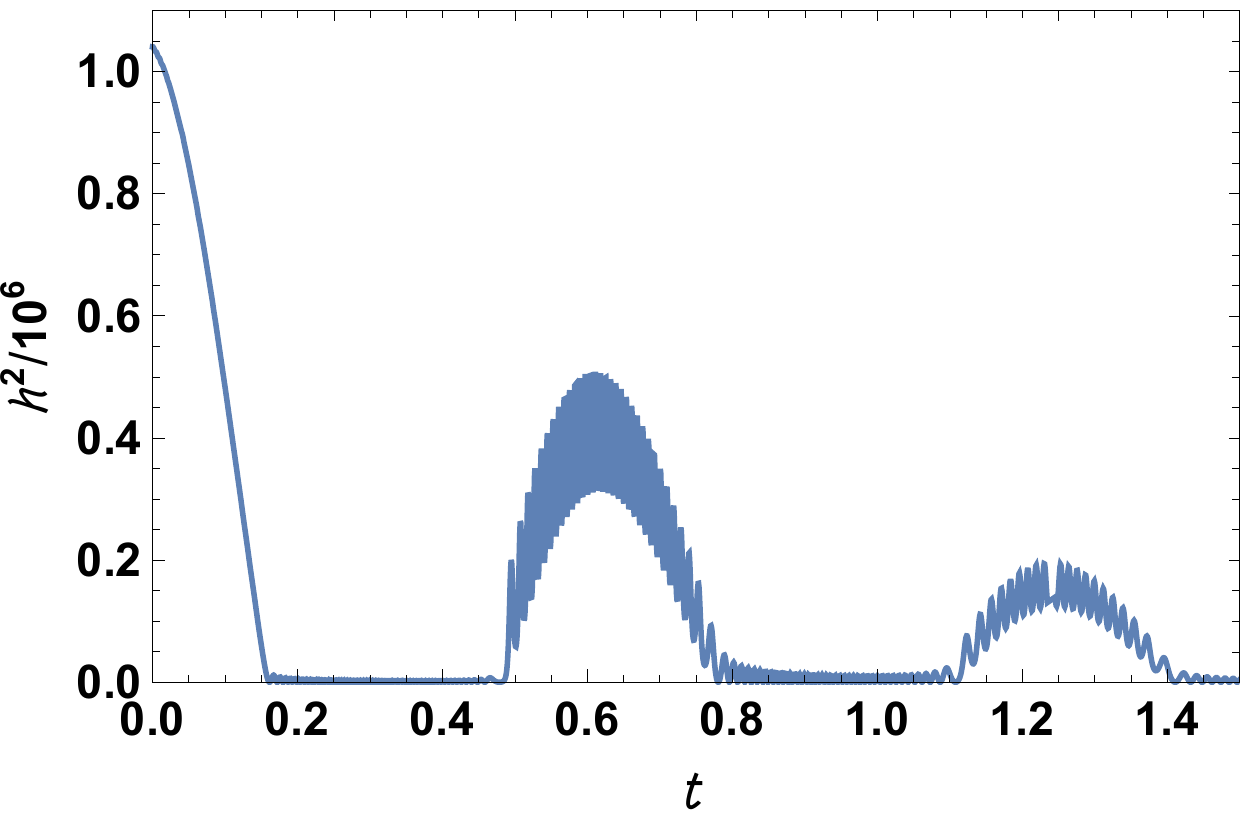} \quad  \includegraphics[width=0.43\textwidth]{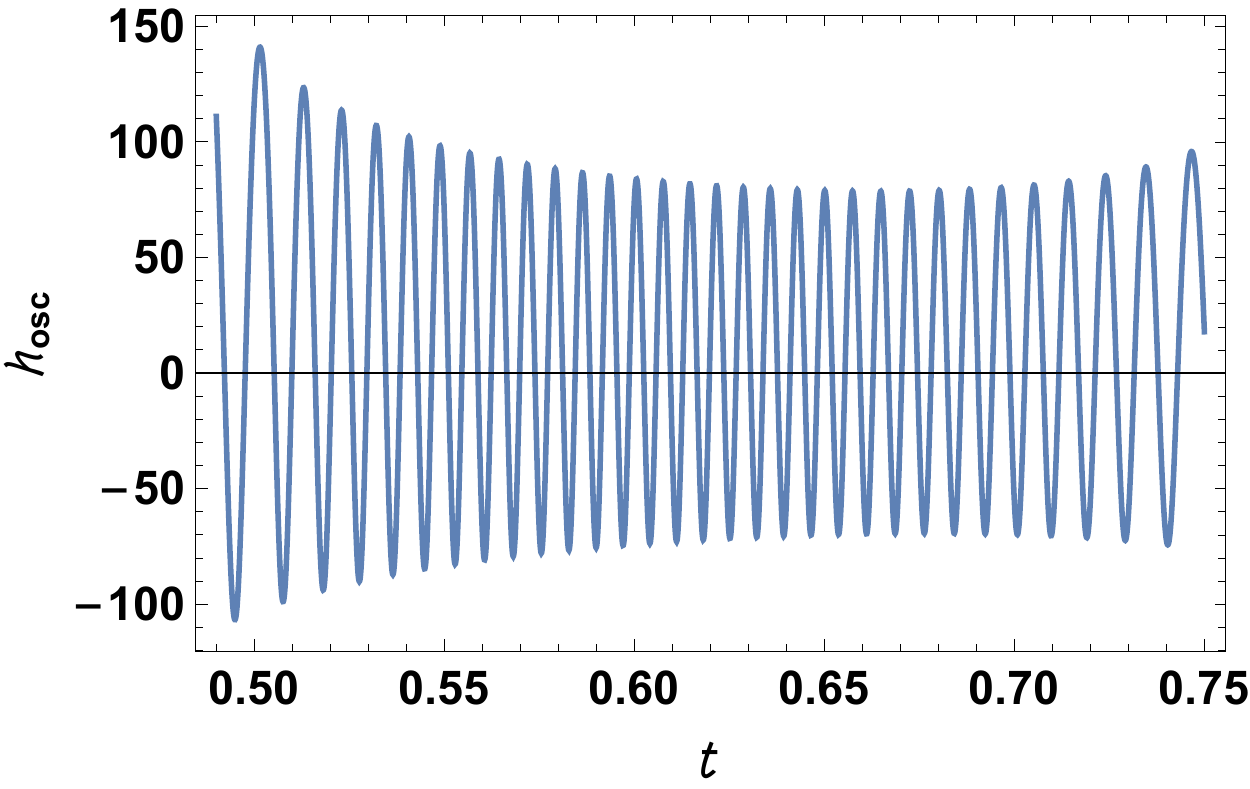} 
    \caption{\label{fig:hsquare}
Left: $h^2$ as a function of $t$. Right: $h_{\rm osc}$ as a function of $t$ in the symmetry breaking phase. We fix $M= 1020 H, m_\chi = 10 H, m_h=2H, \lambda =1$.  
    }
\end{figure}

The inflaton couples to $h^2$. Given the energy hierarchy we consider, $|h_\mathrm{osc}| \ll |h_\mathrm{vev}|$ in the broken phase. An explanation could be found in Appendix~\ref{app:higgs}. In the symmetric phase $h^2 = h_\mathrm{osc}^2$ and in the broken phase $h^2 \simeq h_\mathrm{vev}^2 + 2h_\mathrm{vev} h_\mathrm{osc}$. So it is dominantly the broken phases that could modify the inflaton two-point function. For the benchmark, $h^2$ and $h_{\rm osc}$ as a function of time are presented in Fig.~\ref{fig:hsquare}. We assume that modulus starts from the negative side and Higgs is initially in the vacuum. Yet as one could see from Fig.~\ref{fig:hsquare}, the fast oscillations around the instant minimum are generated after the Higgs oscillates into the different phase. (Similar time dependence of Higgs phases has been previously observed in a model with direct inflaton--Higgs couplings \cite{He:2018gyf,He:2018thesis}.)

In the symmetric phase ($m_\mathrm{eff}^2>0$), $h_\mathrm{vev}=0$. In the broken phase ($m_\mathrm{eff}^2<0$),
\begin{align}
    h_\mathrm{vev}(t) = \sqrt{\frac{-m_\mathrm{eff}^2(t)}{\lambda} }~.
    \label{eq:hvev}
\end{align}
Using WKB approximation, we find that $h_{\rm osc}(t)$ could be written as 
\begin{align}\label{eq:hosc1}
  &  h_\mathrm{osc}(t) = A(t) e^{i\theta(t)}~,
    \qquad
    \theta(t) \equiv \sqrt{2} \int^t |m_\mathrm{eff}(t')| dt'~, \quad m_{\rm eff}^2 < 0,  \\
   &     h_\mathrm{osc}(t) = \tilde{A}(t) e^{i\tilde{\theta}(t)}~,
    \qquad
    \tilde{\theta}(t) \equiv  \int^t |m_\mathrm{eff}(t')| dt'~, \quad \quad \; \; m_{\rm eff}^2 > 0,
     \label{eq:hosc} 
\end{align}
where $A(t), \tilde{A}(t)$ are slowly varying functions compared to the $m_\mathrm{eff}$ scale. From the equation above, one could see that $h_{\mathrm{osc}}(t)$ oscillates with a frequency $\sim M$. 

In summary, there are three types of oscillations with different time scales in the Higgs evolution: 
\begin{itemize}
\item Slow oscillation of $h_{\rm vev}$ with a period $T_{\rm vev} \sim 2\pi/m_\chi$. 
\item Fast oscillation of $h_{\rm osc}$ with a period $T_{\rm osc} \sim 2 \pi/M$. 
\item Intermediate oscillation of $h_{\rm osc}$ with a period $T_{\rm tran} \sim 2 \pi/\left(M^2 m_\chi\right)^{1/3}$ near the transition from symmetric to symmetry breaking phase or vice versa. An explanation of this period could be found in Appendix~\ref{app:higgs}. 
\end{itemize}

\section{Modification of primordial spectrum}
\label{sec:primordialspectrum}

\subsection{The fluctuation of the inflaton}

In Fourier space, the fluctuation of the inflaton $\delta\phi_\mathbf{k}(\tau)$ has the action
\begin{align}
    S = \int d\tau \frac{d^3k}{(2\pi)^3} ~ \frac{1}{2H^2\tau^2} \left ( 1+\frac{y}{\Lambda^2} h^2 \right )
    \left ( \delta\phi_{\mathbf{k}}'\delta\phi_{-\mathbf{k}}' -k^2 \delta\phi_{\mathbf{k}}\delta\phi_{-\mathbf{k}} \right )~,
\end{align}
where the conformal time $\tau \simeq -1/(aH)$ is used and a prime denotes a derivative with respect to the conformal time. Let
\begin{align}
    \alpha \equiv \frac{1}{H^2\tau^2} \left ( 1+\frac{y}{\Lambda^2}h^2  \right )~,
    \qquad \varphi = \alpha^{\frac{1}{2} }\delta\phi~.
\end{align}
Then the canonically normalized field $\varphi$ has an action
\begin{align}
    S = \int d\tau \frac{d^3k}{(2\pi)^3} \frac{1}{2} 
    \left  \{
        \varphi_{\mathbf{k}}'\varphi_{-\mathbf{k}}' - 
        \left  [
            k^2 - \frac{1}{4} \left ( \frac{\alpha'}{\alpha}  \right )^2 
            - \frac{1}{2} \partial_\tau \left ( \frac{\alpha'}{\alpha}  \right )
        \right ] \varphi_{\mathbf{k}}\varphi_{-\mathbf{k}}
    \right \} ~.
\end{align}
The field $\varphi_\mathbf{k}$ is quantized as
\begin{align}
    \varphi_\mathbf{k}(\tau) = u_k(\tau) a_\mathbf{k} + u^*_k(\tau) a^\dagger_{-\mathbf{k}}~,
    \qquad
    [a_\mathbf{k},a_\mathbf{p}]=0~,
    \qquad
    [a_\mathbf{k},a^\dagger_\mathbf{p}]=(2\pi)^3 \delta^3(\mathbf{k}-\mathbf{p})~,
\end{align}
with the mode function $u_k$ satisfying the classical equation of motion
\begin{align}\label{eq:ueom}
    u_k'' + \left  [
        k^2 - \frac{1}{4} \left ( \frac{\alpha'}{\alpha}  \right ) ^2
        - \frac{1}{2} \partial_\tau \left ( \frac{\alpha'}{\alpha}  \right )
    \right ] u_k = 0~,
\end{align}
and the corresponding Bunch-Davies initial condition is
\begin{align}
    u_k(\tau\rightarrow -\infty) = \frac{1}{\sqrt{2k}} e^{-i k \tau}~. 
    \label{eq:ini}
\end{align}

The power spectrum at late times ($\tau\rightarrow 0^-$) can be calculated as follows: the curvature perturbation $\zeta$ in a spatially flat gauge is 
\begin{align}
    \zeta = - \frac{H}{\dot\phi}\delta\phi ~.
\end{align}
The power spectrum $P_\zeta(k)$ is defined from
\begin{align}
    \left\langle \zeta_\mathbf{k} \zeta_\mathbf{p} \right\rangle
    = (2\pi)^3 \delta^3 (\mathbf{k}+\mathbf{p}) \frac{2\pi^2}{k^3} P_\zeta(k)~. 
\end{align}
Thus, the power spectrum can be calculated using
\begin{align}
    P_\zeta(k) = P_\zeta^{(0)} \times \lim_{\tau\rightarrow 0} \left[ 2k^3\tau^2|u_k(\tau)|^2 \right] ~,
    \label{eq:twopoint}
\end{align}
where $P_\zeta^{(0)}$ is the power spectrum of minimal single field inflation without the coupling to the Higgs: $P_\zeta^{(0)}=H^4/(4\pi^2\dot\phi^2)$. Here we are assuming an approximate factorization, because the typical deviations from scale invariance expected during inflation (necessary for instance to fit the observation that $n_s \neq 1$) arise on longer time scales than the rapid oscillations that we are studying.

\subsection{Approximate analytical solution of the inflaton wavefunction}

In the limit of a small coupling $y$ between inflaton and Higgs, we can obtain an approximate analytical solution to the inflaton equation of motion in Eq.~\eqref{eq:ueom}. Eq.~\eqref{eq:ueom} can be approximately written as
\begin{align}\label{eq:u-eom-approx}
    u_k'' + \left  [ k^2 - \frac{2}{\tau^2} \right ] u_k =  \frac{y \partial_\tau^2(h^2)}{2\Lambda^2}  u_k ~.
\end{align}
We drop a first derivative term of $h^2$, which is subdominant compared to the second order derivative term on the right hand side of the equation above since each time derivative brings down the effective mass of $h$, which is of order ${\cal O}(M)$ and is much greater than the Hubble scale.

When $\frac{y \overline{\partial_\tau^2(h^2)}}{2\Lambda^2} \ll k^2$, this equation can be solved perturbatively. We write $u_k \simeq u_k^{(0)}+u_k^{(1)}+\cdots$.
In the absence of the perturbation term, the leading solution is the Bunch-Davies mode function,  
\begin{align}
    u_k^{(0)} = 
    \frac{1}{\sqrt{2k}} 
    \left(
        1 - \frac{i}{k\tau}
    \right) e^{-ik\tau}~.
    \label{eq:leadingsol}
\end{align}
The leading perturbation, $u_k^{(1)}$ satisfies the equation 
\begin{align}
u_k^{(1)''}+ \left  [ k^2 - \frac{2}{\tau^2} \right ] u_k^{(1)}=  \frac{y \partial_\tau^2(h^2)}{2\Lambda^2}  u_k^{(0)} ~.
\end{align}
Since $u_k$ has to satisfy the Bunch-Davies initial condition, the initial conditions for $u_k^{(1)}$ is 
\begin{align}
u_k^{(1)} (\tau \to - \infty) = 0,  \quad u_k^{(1)\prime} (\tau \to - \infty) = 0
\label{eq:uk1ini}
\end{align}

Using the Green function method (or variation of parameters), we find that 
\begin{align}
u_k^{(1)}(\tau) &= \frac{i y}{2\Lambda^2 (2 k)^{3/2}}\left\{- e^{i k \tau} \left(1+ \frac{i}{k \tau}\right) \int_{-\infty}^\tau d\eta \; e^{-2ik\eta}\left(1- \frac{i}{k \eta}\right)^2 \partial^2_\eta (h^2(\eta))  \right. \nonumber \\
&\quad\quad\quad\quad\quad\quad
\left. + \frac{e^{-i k \tau}}{k^2} \left(1-\frac{i}{k\tau}\right) \int_{-\infty}^\tau d\eta \;  \left(\frac{1}{\eta^2}+k^2\right)\partial^2_\eta \left(h^2(\eta)\right) \right\} \nonumber \\
&\underset{\tau \to 0}{=} - \frac{1}{k\tau} \frac{y}{2\Lambda^2 (2 k)^{3/2}}   \int_{-\infty}^0 d\eta \; \Bigg[\partial_\eta \left(\left(1- \frac{i}{k \eta}\right)^2e^{-2ik\eta}\right) - \frac{2}{k^2\eta^3}\Bigg]  \partial_\eta (h^2(\eta))
\label{eq:approxlatetime}
\end{align}
is the solution obeying the initial conditions in Eq.~\eqref{eq:uk1ini}. We have used integration by parts to reduce the second derivative on $h^2$ to first order, and dropped a boundary term at $\tau \rightarrow -\infty$, since the evolution of Higgs has not started at the initial time. 
In the limit of early times with $- k \tau \gg 1$, the solution above reduces to  
\begin{align}\label{eq:higgs-sol-1}
    u_k^{(1)} \approx  \frac{y e^{ik\tau}}{2\Lambda^2 \sqrt{2k}} 
    \int_{-\infty}^\tau d\eta ~ e^{-2ik\eta}\partial_\eta [h^2(\eta)]
    \approx 
    \frac{i y \sqrt {2k} e^{ik\tau}}{2\Lambda^2} 
    \int_{-\infty}^\tau d\eta ~ e^{-2ik\eta} h^2(\eta)~.
\end{align}
Note that this is simply the Fourier transform of $h^2$.
 
Below we will use the approximate perturbative solution to derive corrections to the primordial inflaton spectrum and compare them with the exact numerical solutions. As we argued in Sec.~\ref{sec:higgsprofile}, it is dominantly the broken phases that affect the inflaton spectrum given their larger $h^2$. We thus focus on the broken phases below. In a broken phase, $h^2(t) \approx h_{\rm vev}^2 + 2 h_{\rm vev} h_{\rm osc}$. Since the two terms have different oscillation frequencies, they affect different $k$ ranges. We will go through each of them separately in the discussion below. 

\subsubsection{$h_{\rm vev}^2$ contribution}
\label{sec:hvev}
We start with the contribution to the inflaton two-point function from $h_{\rm vev}^2$. Note that $h_{\rm vev}$ is not a full cosine function since it is zero in the symmetry preserving phase. Taking into account that the Higgs starts from the symmetry breaking side with $\chi_0 = -f$ at $t=0$, we have
\begin{equation}
h^2_{\rm vev}(t) = \begin{cases}
          \frac{M^2}{\lambda}, & t < 0 \\
           \frac{M^2}{\lambda} \cos(m_\chi t), & 0 \leq t \leq \frac{\pi}{2m_\chi} \\
           \frac{M^2}{\lambda} e^{-\frac{3(n+1)\pi H}{m_\chi}} \cos(m_\chi t), & \left(\frac{3}{2} + 2n \right)\frac{\pi}{m_\chi} \leq t \leq  \left(\frac{5}{2} + 2n \right)\frac{\pi}{m_\chi}, \quad n \in \mathbb{Z}_{\geq 0},
       \end{cases}    
\end{equation} 
where we ignore the $-m_h^2$ terms and approximate the amplitude redshift in each symmetry breaking phase by the one at the maximally symmetry breaking point in that phase.\footnote{In a complete model, the modulus will slowly evolve before it begins oscillating, and $h_{\rm vev}(t)$ will have nontrivial evolution at $t < 0$. Because we are interested in the effects of an oscillating modulus, however, we use the simplifying ansatz that the Higgs does not evolve until $t = 0$.}

To implement the conformal time integration in Eq.~\eqref{eq:approxlatetime}, we write $h_{\rm vev}^2(t)$ in terms of its Fourier transform $\tilde{h}(\omega)$:
\begin{align}
h^2_{\rm vev}(t) &= \int_{-\infty}^{\infty} \tilde{h}(\omega) e^{2\pi i \omega t} d \omega, \nonumber \\
 \tilde{h}(\omega) &= \frac{M^2}{\lambda} \frac{m_\chi}{m_\chi^2-4\pi^2\omega^2}  \left[  i  \frac{m_\chi}{2\pi \omega}+ e^{-\frac{i \pi^2 \omega}{m_\chi}} + \sum_{n} e^{-\frac{3(n+1)\pi H}{m_\chi}}\left(e^{-\frac{(3+4n) i \pi^2 \omega}{m_\chi}} +e^{-\frac{(5+4n) i \pi^2 \omega}{m_\chi}}  \right)\right],
\label{eq:Fourier}
\end{align} 
where the physical time $t$ is related to the conformal time $\eta$ as 
\begin{align}
t = -\frac{1}{H} \ln \frac{\eta}{\eta_0}=-\frac{1}{H} \ln \left(-H\eta \right), \quad \eta_0=- \frac{1}{a_0H} = -\frac{1}{H}.
\end{align}
Here the subscript $0$ indicates the time that the modulus starts to oscillate, since it corresponds to the initial time $t=0$. We will use the convention $a_0=1$. 

\begin{figure}[htbp] \centering
    \includegraphics[width=0.55\textwidth]{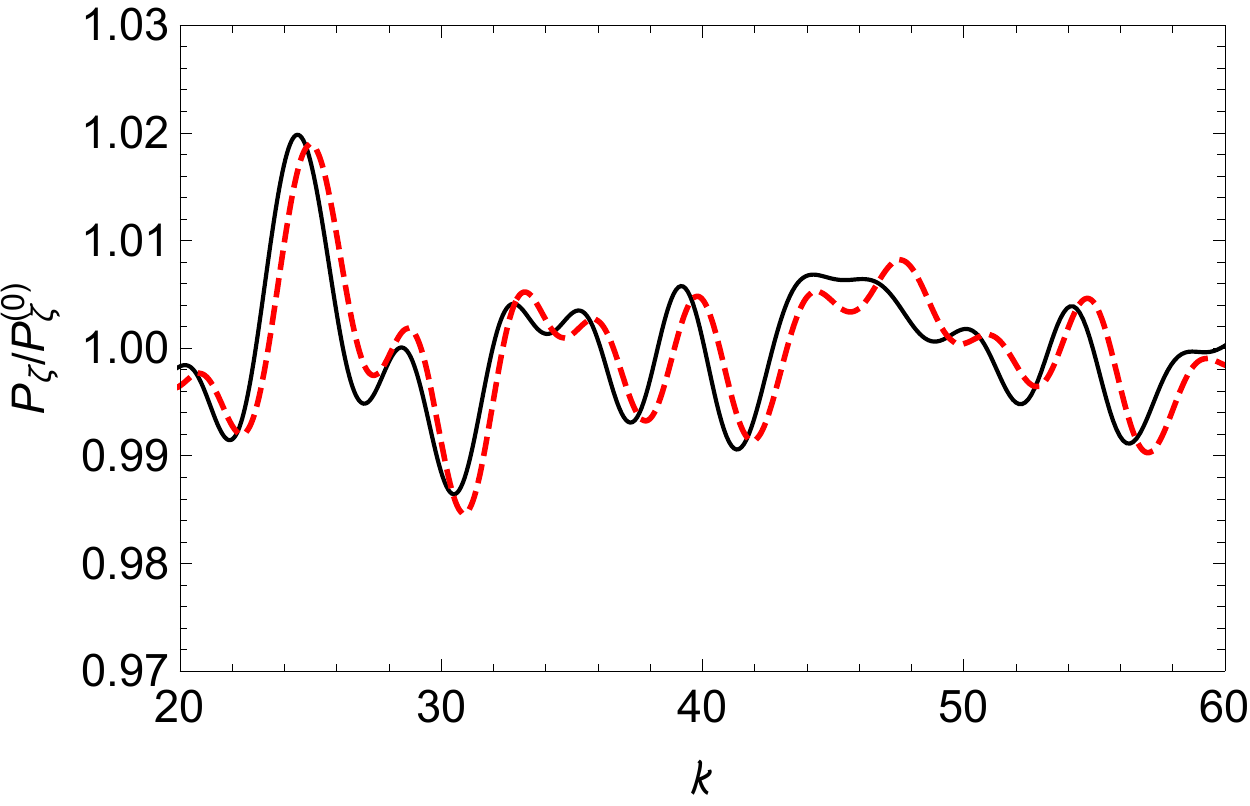} 
    \caption{Correction to the primordial spectrum in the low $k$ range. $k$ is in unit of $H$. We fix $\Lambda = 6000 H, M= 1020 H, m_\chi = 10 H, m_h=2H, \lambda =1$. Black: full numerical result; red dashed: analytical estimate using Eq.~\eqref{eq:lowkcor} including terms  up to $n = 5$. \label{fig:lowk}
    }
\end{figure}

Plugging the Fourier expression of $h^2_{\rm vev}$ into the last equality of Eq.~\eqref{eq:approxlatetime} and integrating first $\eta$ and then $\omega$ using saddle point approximations (Appendix~\ref{app:saddle}), we find that the induced correction to the inflaton wavefunction at $\tau \to 0$ is, for $k > m_\chi$, 
\begin{align}
(k\tau)u_{k; {\rm vev}}^{(1)} 
&\approx \frac{y}{2\Lambda^2(2k)^{3/2}}\frac{M^2m_\chi}{\lambda}  \vast\{\frac{4k^2}{4k^2-m_\chi^2}e^{i\frac{2k}{H}}\left(\frac{i m_\chi}{2k} + e^{-i\frac{\pi k}{m_\chi}}\right) \nonumber \\
&\quad \quad \quad\quad\quad\quad\quad+\sum_n \frac{8k^2 e^{-\frac{5(n+1)\pi H}{m_\chi}}}{4k^2e^{-\frac{4(n+1)\pi H}{m_\chi}}-m_\chi^2}e^{i\frac{2k}{H}e^{-\frac{2(n+1)\pi H}{m_\chi}}}\cos\left(\frac{\pi k e^{-\frac{2(n+1)\pi H}{m_\chi}}}{m_\chi}\right)\vast\}.
 \end{align}
More details of the derivation could be found in Appendix~\ref{app:vev}. 
Then its contribution to the modified two-point function is 
\begin{align}
\frac{\delta P_\zeta(k)}{P_\zeta^{(0)}} &\approx \lim_{\tau \to 0} \left[2k^3\tau^2 \left(u_k^{(0)*} u_k^{(1)} + u_k^{(0)} u_k^{(1)*}\right)\right]  \nonumber \\
&=-\frac{y m_\chi M^2 }{2 \lambda \Lambda^2 k}\Biggl\{ \frac{4k^2}{4k^2-m_\chi^2} \left[\sin\left(\frac{2k}{H}\right) \cos\left(\frac{\pi k}{m_\chi}\right) + \cos\left(\frac{2k}{H}\right)\left(\frac{m_\chi}{2k} - \sin\left(\frac{\pi k}{m_\chi}\right)\right)\right] \nonumber \\
& \quad \quad \quad \quad \quad+ \sum_{n=0} \frac{8k^2e^{-\frac{5(n+1)\pi H}{m_\chi}}}{4k^2e^{-\frac{4(n+1)\pi H}{m_\chi}} - m_\chi^2} \sin\left(\frac{2k}{H}e^{-\frac{2(n+1)\pi H}{m_\chi}}\right)\cos\left(\frac{\pi k}{m_\chi}e^{-\frac{2(n+1)\pi H}{m_\chi}}\right)\Biggr\}.
\label{eq:lowkcor}
\end{align} 
Since the oscillation amplitudes redshift, the first few cycles contribute dominantly to the primordial spectrum. Then based on the equation above, the correction in the low $k$ region is a superposition of periodic functions with similar frequencies around $\pi H$, of which the differences are exponentially suppressed since $m_\chi \gg H$. This will lead to oscillations in the low $k$ range with an approximately constant frequency but varying amplitude. This is shown in Fig.~\ref{fig:lowk}.

Note that the spectrum with ``beats" is a result of the Higgs profile being a piecewise cosine function. If $h^2_{\rm vev}$ is a full cosine function such as $\cos(m_\chi t)$, the correction to the primordial spectrum would take the form of $\sin(C \log k)$, as studied in the context of the classical primordial clock~\cite{Chen:2011zf}.

\subsubsection{$h_{\rm vev}h_{\rm osc}$ contribution} 
\label{sec:hvevhosc}
Now let's consider the fast oscillating $h_{\rm vev}h_{\rm osc}$'s contribution to the primordial spectrum.
Combining Eq.~\eqref{eq:hvev}, \eqref{eq:hosc1} and Eq.~\eqref{eq:approxlatetime}, we have at $\tau \to 0$,
\begin{align}
    u_{k;{\rm osc}}^{(1)} \simeq \frac{y}{\Lambda^2 (2k)^{3/2} (k\tau)} 
    \int_{-\infty}^0 d\eta ~ e^{-2ik\eta + \sqrt{2}i \int^t |m_\mathrm{eff}(t')|dt'}
    \sqrt{\frac{-m_\mathrm{eff}^2(t(\eta))}{\lambda} } A(t(\eta)) f(\eta)~,
\end{align}
where $f(\eta) = \partial_{\eta}^2 \left[\left(1- \frac{i}{k\eta}\right)^2 e^{-2ik\eta}\right] e^{2ik\eta}$, and the relatively slowly varying function $A(t)$ is estimated in \eqref{eq:aestimate} in the appendix.
At resonance, the condition of stationary phase gives
\begin{align}\label{eq:resonance-condition}
    \frac{k}{a(t)} = \frac{|m_\mathrm{eff}(t)|}{\sqrt{2}} ~.
\end{align}
Since $|m_\mathrm{eff}|$ is oscillatory, the resonance happens many times for each $k$, each time with an almost random phase. The amplitude of the resonance also decays at time scales comparable to $1/H$. As an order-of-magnitude estimate, it makes sense to approximate the correction to the wavefunction summing over all resonances, for each $k$ mode:
\begin{align}
    u_{k;\rm{osc}}^{(1)}
    \sim
    \sqrt{\mbox{(number of resonances in } 1/H \mbox{ time)} } \times
    \mbox{(the contribution from the first resonance)}~.
\end{align}
The square root takes into account the randomness of each resonance phase. 
Here, the number of resonances per Hubble time is approximately $m_\chi / (\pi H)$.

\begin{figure}[htbp] \centering
    \includegraphics[width=0.6\textwidth]{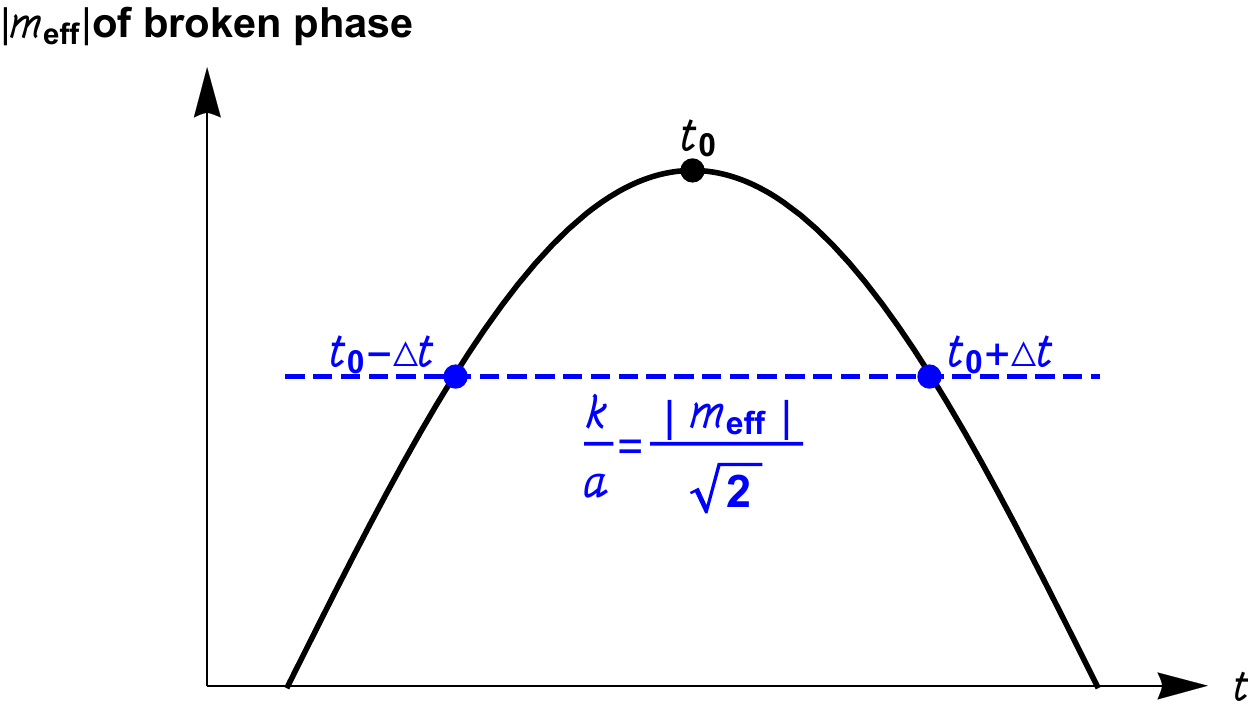}
    \caption{\label{fig:phase_diff}
        Two resonances around a maximally symmetry broken point. 
    }
\end{figure}

To obtain a crude yet relatively clean analytical understanding of the resonance contributions, we study the behavior near a (locally) maximally symmetry broken point $t_0$. This is illustrated in Fig.~\ref{fig:phase_diff}. Below we take $H\rightarrow 0$. In this limit, $a\rightarrow 1$ and $\tau \rightarrow - \frac{1}{H} + t + \mathcal{O}(H)$. We also take $m_h\rightarrow 0$. 
For a given $k$, the resonance condition \eqref{eq:resonance-condition} can be written as
\begin{align}
    \Delta t = \frac{1}{m_\chi} \arccos \left [ \frac{2k^2}{M^2} \frac{f}{-\chi(t_0)}  \right ]~, 
    \label{eq:deltat}
\end{align}
In a single symmetry breaking phase, there are two resonances for a fixed $k$ as shown in Fig.~\ref{fig:phase_diff}. 

Both resonances contribute to the correction of the inflaton wavefunction 
\begin{align}
u_{k;{\rm osc}}^{(1)} = \frac{y}{\Lambda^2 (2k)^{3/2} (k\tau) } \sum_{t_r} \left.\left [ \sqrt{\frac{-m_\mathrm{eff}^2(t)}{\lambda} } A(t) f(\eta) 
\sqrt{\frac{\sqrt{2}\pi i |m_\mathrm{eff}|}{2k^2\left(H+\frac{d \log(-m_\mathrm{eff}(t))}{dt}\right)}}
\right] \right|_{t=t_r} e^{i \theta_r},
\end{align}
where the sum runs over the two resonances at time $t_r$'s and the resonance phases are given by  
\begin{align}
    \theta_r &= -2k \left.\eta\right|_{t=t_r} + \sqrt{2} \int_{t_0}^{t_r} |m_\mathrm{eff}(t')| dt' \nonumber \\
&= 2 \frac{k}{H}  - 2 k  t_r + \sqrt{2} \int_{t_0}^{t_r} |m_\mathrm{eff}(t')| dt'   \nonumber \\
&\approx 2 \frac{k}{H}  - 2 k  \left(t_0  \pm \Delta t\right) + \sqrt{\frac{2}{\lambda}} M e^{-\frac{3Ht_0}{4}} \int_{t_0}^{t_r}\sqrt{\cos \left(m_\chi t' \right)}  dt'   \nonumber \\
&\approx 2 \frac{k}{H} - 2k t_0 \mp \frac{2 k}{m_\chi}  \arccos \left ( \frac{2k^2}{M^2}  \frac{f}{-\chi(t_0)} \right) + \sqrt{\frac{2}{\lambda}} \frac{2M}{m_\chi} e^{-\frac{3Ht_0}{4}}\mathrm{E} 
    \left ( \left. \frac{1}{2} m_\chi (t_r-t_0) \right| 2 \right )
    \label{eq:phase}
\end{align}
where we used the approximation that when $H \to 0$, the conformal time $\eta \approx -\frac{1}{H} + t$. The resonance time $t_r$ takes one of the two values $t_0 \pm \Delta t$ for the first and second resonance, respectively. In the last line above, we use Eq.~\eqref{eq:deltat}. $\mathrm{E}\left(x|2\right)$ is the elliptic integral of the second kind and is sub-dominant in the phase when $M \lesssim 10^3 H$.

\begin{figure}[h] \centering
    \includegraphics[width=0.6\textwidth]{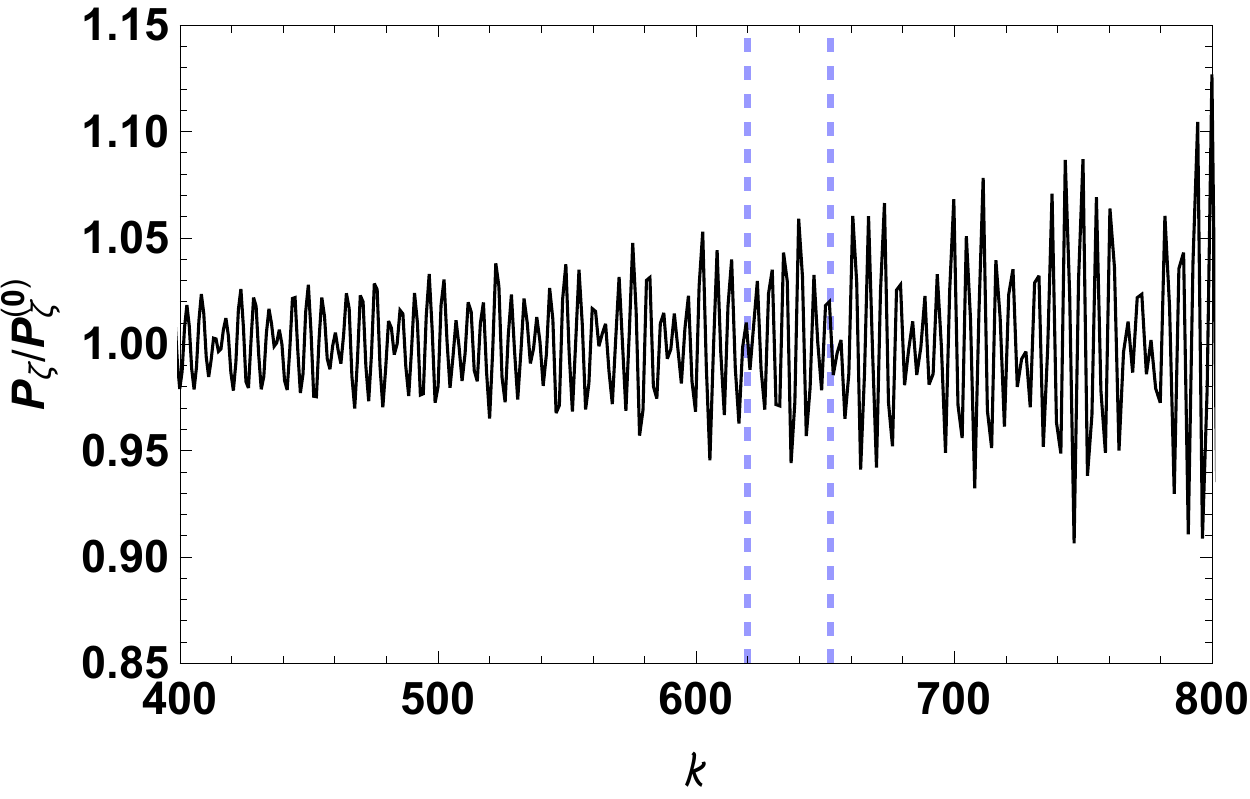} 
    \caption{\label{fig:highk} The ratio $P_\zeta/P_\zeta^{(0)}$ in the high-$k$ range from solving the system numerically. $k$ is in unit of $H$. We fix $\Lambda =6 \times 10^3 H, M= 1020 H, m_\chi = 10 H, m_h=2H, \lambda =1$.  To clarify what we mean by ``k-wavepacket,'' we have marked one between dashed blue lines.
    }
\end{figure}

Numerically, we observe an interesting repeated ``k-wavepacket" feature with two oscillation frequencies in the $k$-space in the correction to the two-point function at large $k$. This is shown in Fig.~\ref{fig:highk}. From the analytic estimate of the phases at resonance in Eq.~\eqref{eq:phase}, we could understand the origins of the two frequencies as follows: 
\begin{itemize}
    \item The large frequency (fast oscillation in $k$-space): since we consider $k\gg H$, the leading oscillation is proportional to $\cos (2k/H)$. In every $\Delta k/H = 10$ range, there should be about 3 peaks, which roughly agrees with the full numerical result in Fig.~\ref{fig:highk}. 
        \item The small frequency (slow modulation in $k$-space): the slow modulation gives the envelope of the fast oscillations. In a single symmetry breaking phase, the two resonances with phases given by Eq.~\eqref{eq:phase} could partially cancel each other at certain $k$'s. More specifically, the common phase $2k/H$, which leads to fast oscillations, cancels out. On the other hand, the terms with opposite signs, $\pm \frac{2 k}{m_\chi}  \arccos \left ( \frac{2k^2}{M^2}  \frac{f}{-\chi(t_0)} \right)$ contribute to the slow modulation. Approximating $\arccos \left ( \frac{2k^2}{M^2}  \frac{f}{-\chi(t_0)} \right)$ by one (in general, this is an order one phase when $k \lesssim M$), the slow oscillation could be approximated as $\cos(2 k / m_\chi)$. For the benchmark we choose, the analytic argument indicates that the slow oscillation period is set by $\pi m_\chi$, which roughly agrees with the numerical result, as shown in Fig.~\ref{fig:highk}.
\end{itemize}

\subsubsection{Summary of signatures on the primordial power spectrum} 
\label{sec:summaryofsignatures}
Here we summarize the observable signatures discussed in this section, and relate them to the parameters of the model. 

\begin{enumerate}
    \item Oscillatory signal from $h_\mathrm{vev}^2$:
    \begin{enumerate}
        \item \label{item:fourier} The signal is the Fourier transform of the Higgs symmetric and breaking phases, which appears as piecewise half cosine functions.
        \item \label{item:mchi} The position of the oscillation in the power spectrum is $k\sim m_\chi$.
        \item \label{item:mch1} The oscillation pattern is a superposition of periodic functions with similar frequencies in the perturbative limit. Including the first two oscillations, it is $e^{\frac{\pi H}{2m_\chi}} \sin\left(\frac{2k}{H}e^{-\frac{\pi H}{2m_\chi}}\right) + e^{-\frac{3\pi H}{2m_\chi}}\sin\left(\frac{2k}{H}e^{-\frac{3\pi H}{2m_\chi}}\right)+ e^{-\frac{\pi H}{2m_\chi}}\sin\left(\frac{2k}{H}e^{-\frac{5\pi H}{2m_\chi}}\right)$. 
        \item \label{item:d1} The oscillation amplitude is of order $\Delta_\mathrm{vev}\equiv \frac{\delta P_\zeta}{P^0_\zeta}\sim \frac{yM^2m_\chi}{\lambda\Lambda^2 k} $.
    \end{enumerate}
    \item Oscillatory signal from $h_\mathrm{vev} h_\mathrm{osc}$:
    \begin{enumerate}
        \item The signal records the resonances of each sub-horizon perturbation mode with the time-varying Higgs mass.
        \item \label{item:mh1} The position of the oscillation in the power spectrum is $k\sim M$.
        \item The oscillation pattern is in the form of k-wavepackets, including:
        \begin{enumerate}
            \item \label{item:kh1} Fast oscillations of the shape $\cos(2k/H)$ from the resonances.
            \item \label{item:mh2} A slow modulation (envelope) of the shape $\cos(2 k/m_\chi)$ (the form is given in general in \eqref{eq:phase}) from the constructive and destructive interferences of the resonances.
        \end{enumerate}
        \item \label{item:d2} The oscillation amplitude is of order $\Delta_\mathrm{osc}\equiv \frac{\delta P_\zeta}{P^0_\zeta}\sim \frac{yM^{13/6}}{\lambda\Lambda^2 m_\chi^{1/6}}$.
                \end{enumerate}
    \item \label{item:kh2} Further, there should be a feature in the power spectrum happening at $k\sim H$, when the modulus oscillation is triggered \cite{Chen:2011zf}. The shape of the feature depends on how the modulus starts to oscillate.
\end{enumerate}

From this summary, in principle, if we had excellent knowledge about the primordial power spectrum, the model parameters could be inferred as follows:

\begin{itemize}
    \item The relation between $k$ and $H$ can be obtained from items \ref{item:kh2} and \ref{item:kh1} of the above enumerated list. After fixing the relation between $k$ and $H$, every feature discussed below can be cast in Hubble units.
    \item The ratio $M/H$ can be obtained from \ref{item:mh1}.
    \item The ratio $m_\chi/H$ can be obtained from \ref{item:mchi} and \ref{item:mch1}.
    \item $\lambda\Lambda^2$ can be obtained from \ref{item:mch1} and \ref{item:d2} once all the other mass scales are determined as above. 
\end{itemize}

A further possible difficulty is that the high-$k$ spectrum proves to be somewhat sensitive to the precise choices of parameter values. Our plots so far have shown a benchmark choice $M = 1020 H$, which we chose as a relatively optimistic case in terms of the clarity of the k-wavepackets and their amplitude for fixed values of $\Lambda$ and other parameters. We show in Appendix \ref{app:paramsensitivity} that other nearby choices of $M$ produce similar features, but in some cases the amplitude is smaller and in other cases the k-wavepacket features are less distinct. This sensitivity seems to arise from the precise timing of where the Higgs field oscillation $h_{\rm osc}$ lies in its cycle at the moment that the phase switching due to the modulus occurs.

In reality, observation of some of the biggest features listed above would already be a great probe of the physics. Despite having uncertainties, such observations could give us knowledge of the symmetry and symmetry-breaking phases. It will be more challenging to infer the amount of fine-tuning $m_h/M$, since it will involve the fine structure of the transformed, noisy data. Nevertheless, it is interesting to see that such information has been encoded in the sky in the primordial universe. 

\section{CMB observables} 
\label{sec:CMB}
\begin{figure}[h] 
\begin{center}
 \includegraphics[width=0.45\textwidth]{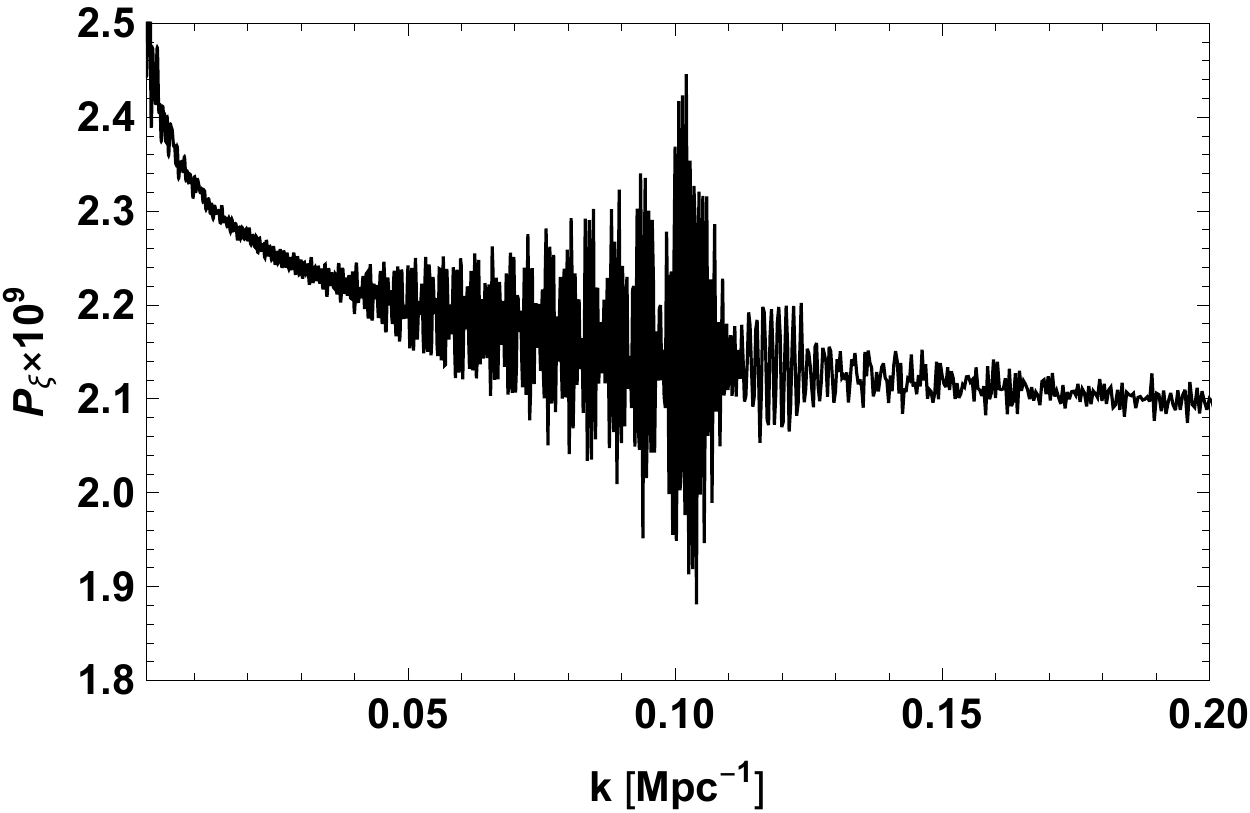}  \includegraphics[width=0.46\textwidth]{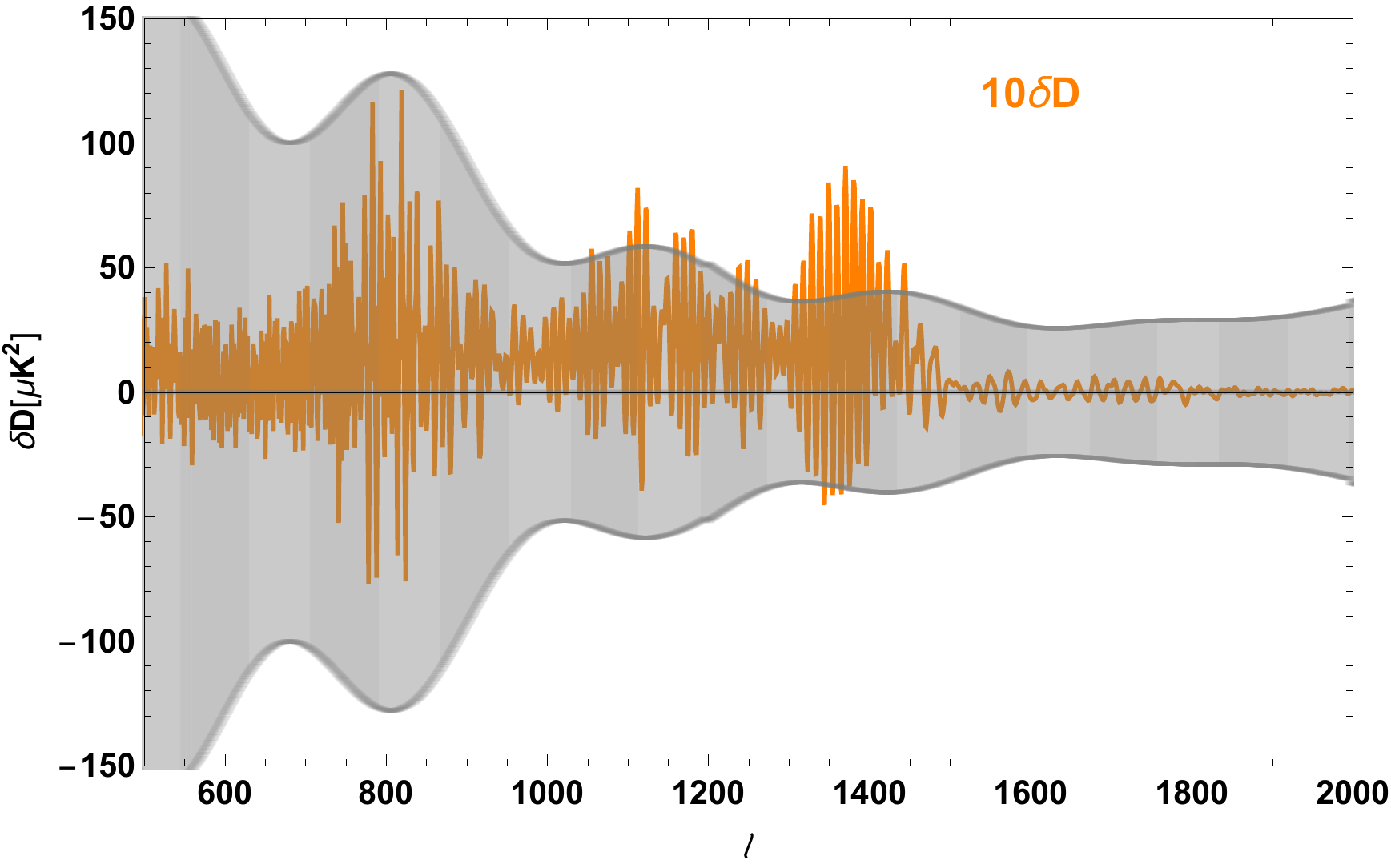}  
\caption{ Left: primordial spectrum resulting from phase transition oscillations fixing $\Lambda = 6000 H, M= 1020 H, m_\chi = 10 H, m_h=2H, \lambda =1$, adding on top of a smooth spectrum with $n_s = 0.9649$. 
Right: the corrections in the CMB temperature spectrum by subtracting that of the inflation model with $n_s = 0.9649$ (right). In plotting, we have multiplied the oscillating correction by a factor of 10. The grey band denotes the unbinned Planck uncertainties, which are $\sim 5\%$ when expressed as fractional uncertainties.
\label{fig:CMB} 
    }
\end{center}
\end{figure}

The oscillation patterns in the primordial spectrum lead to fine structures in the CMB spectrum. Here we focus on the temperature anisotropy spectrum. We feed the primordial spectrum into the CLASS package~\cite{2011arXiv1104.2932L} to compute the CMB temperature spectrum. The result for the benchmark model is presented in Fig.~\ref{fig:CMB}. For the input primordial spectrum, we add the modifications due to phase transition oscillations on top of a smooth spectrum with $n_s = 0.9649$, which is the central value of the spectral index of scalar perturbations determined by Planck temperature, polarization, and lensing data~\cite{Akrami:2018odb}. 

The range of angular multipole moments $\ell$ that could be affected is determined by the comoving wave number compared to the scale of the observable universe. Roughly speaking,
\begin{align}
\ell \sim \frac{k}{H_0} \sim \frac{k}{ 10^{-4} \,{\rm Mpc}^{-1}},
\end{align}
where $H_0$ is the Hubble today. From Fig.~\ref{fig:CMB}, we observe that a 10\% modification in the primordial spectrum leads to a $\sim 1\%$ modification in a wide range of $\ell$'s in the CMB temperature spectrum. The reason for the reduction of the modification in the CMB spectrum is that the temperature harmonic power spectrum is given by a convolution 
\begin{align}
C_\ell \equiv \frac{1}{2\pi^2} \int \frac{dk}{k} \Theta_\ell^2(k) {\cal P}_{\zeta}(k), 
\end{align}
where $\Theta_\ell$ is a transfer function and ${\cal P}_{\zeta}$ is the primordial spectrum. The oscillations in the primordial spectrum are thus smoothed out by the integration, reducing the amplitude.

It is challenging to search for the oscillatory signal of order $1\%$ that we present in this benchmark model, since unbinned Planck data has large error per $\ell$, at order $10\%$ or even larger (though the error bar for binned data is much smaller, the signal is further averaged away). For the search for the signal in Planck data, it remains interesting to further explore two possibilities: {\it i)} Explore the parameter region where the primordial power spectrum gets non-perturbatively large corrections; and {\it ii)} carry out a more careful statistical analysis and template-based search on the Planck unbinned data. These are beyond the scope of this paper.

In the future, the upcoming experiments such as CMB-S4 \cite{Abazajian:2016yjj} will further improve the high-$\ell$ observation and provide better data for searching for this feature. Also, it is interesting to see if the oscillation leaves more observable effects in the large scale structure and the future 21 cm surveys, since they may suffer less from projection effects. Preliminary studies using either CMB or large scale structure on searching for simple $\sin(C \log k)$ oscillation feature in the primordial spectrum could be found in Ref.~\cite{Slosar:2019gvt}.

Here we have assumed that the feature lies at high $\ell$. On the other hand, if the oscillatory feature happened much earlier and appears in the low-$\ell$ CMB multipoles, it would be related to the possible parity asymmetry observation hinted at by WMAP \cite{Bennett:2010jb} and Planck \cite{Schwarz:2015cma}. However, with the large cosmic variance at low $\ell$, we do not expect to get much information about the underlying particle physics even if the oscillatory power spectrum may improve the fitting of data.

\section{Comparisons between different models} 
\label{sec:compare}
One natural question one could ask is that whether the k-wavepacket feature in the primordial spectrum that we find from the phase oscillation model could show up in a different model. In other words, could there be a degeneracy in the model space for the same signal? We survey several different kinds of models and do not find a model which could fully mimic the patterns both in the low $k$ and high $k$ ranges as in the phase oscillation model. Yet it is still possible at the qualitative level that the k-wavepacket feature shows up in a different class of models with different details. One example in the literature is the ``drifting axion monodromy" model~\cite{Flauger:2014ana}, in which the axion inflaton itself oscillates with a drifting period set by moduli fields that it couples to. Below we will present a comparison between a model with no phase transition when Higgs field oscillates, and the phase oscillation model. 

The Lagrangian of the new model is different from Eq.~\eqref{eq:Lagrangian} in the modulus coupling to the Higgs, 
\begin{align}
V \supset -\frac{M^2}{2f^2} \chi^2 h^2.
\end{align}
In this model, as the modulus oscillates, the Higgs is always in the broken phase. We consider two initial conditions for the Higgs field: {\it a)} the Higgs starts close to the origin of the field space and thus on the top of potential hill; {\it b)} the Higgs starts close to one of the minima in the potential. 

\begin{figure}
\begin{center}
    \includegraphics[width=0.6\textwidth]{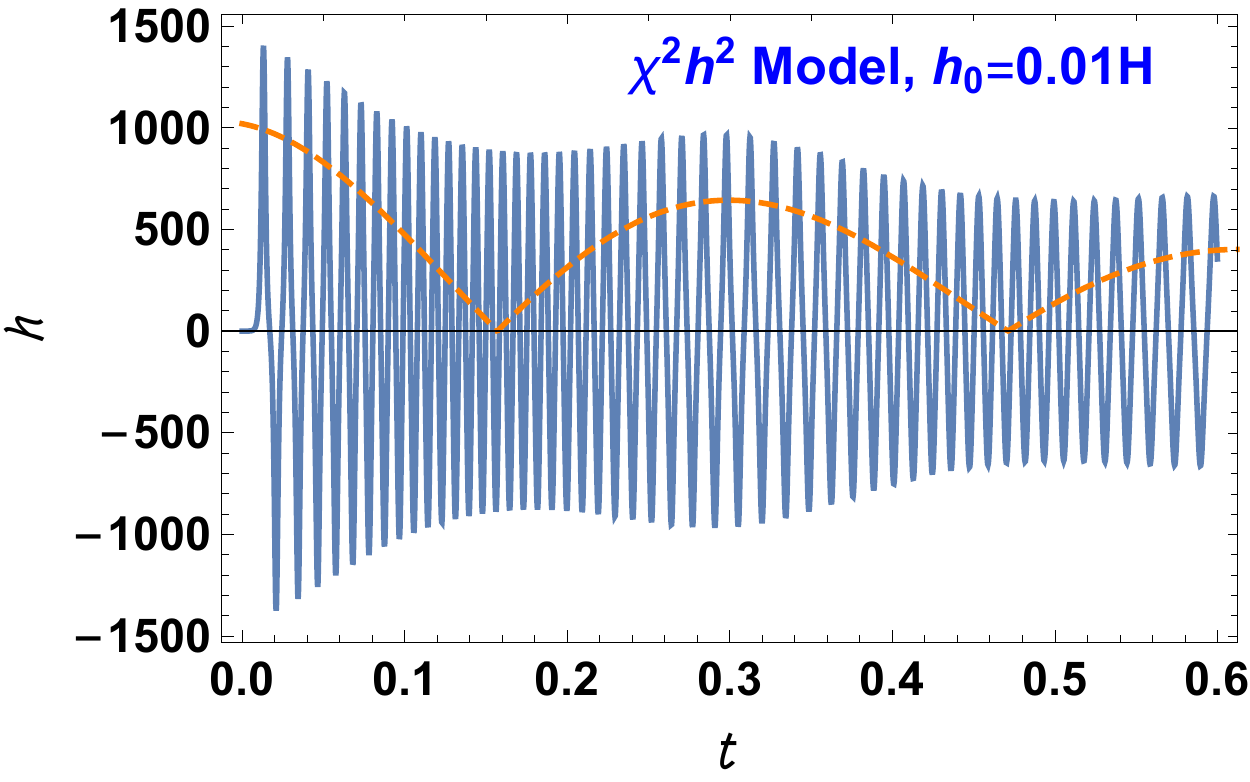} 
    \caption{\label{fig:sloshzoomin} The Higgs evolution in the model where the Higgs is always in the broken phase and Higgs starts near the top of the potential hill. We set $M = 1020H, m_\chi = 10 H, \chi_0 =f$ and $\lambda = 1$. The dashed orange curve is $M |\cos(m_\chi t)|$. 
    }
   \end{center} 
\end{figure}

\begin{figure}[h] 
    \includegraphics[width=0.32\textwidth]{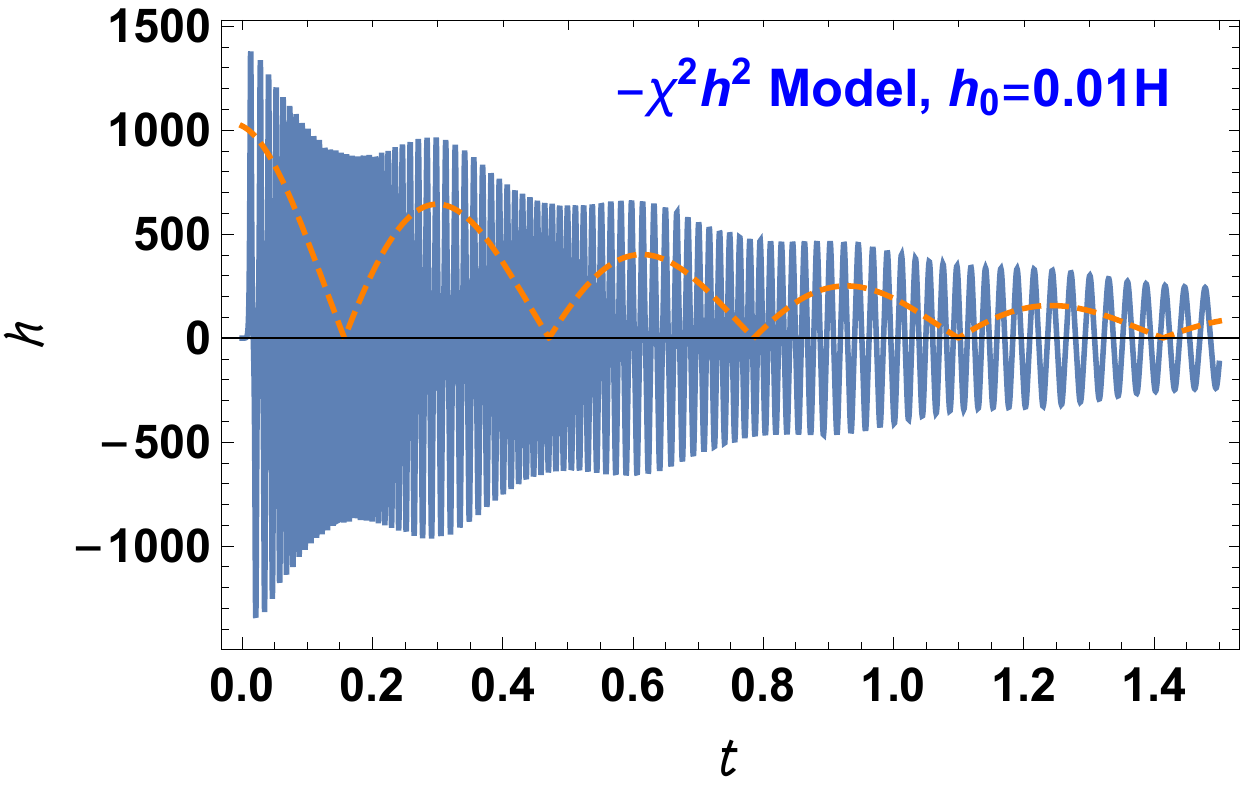} \includegraphics[width=0.32\textwidth]{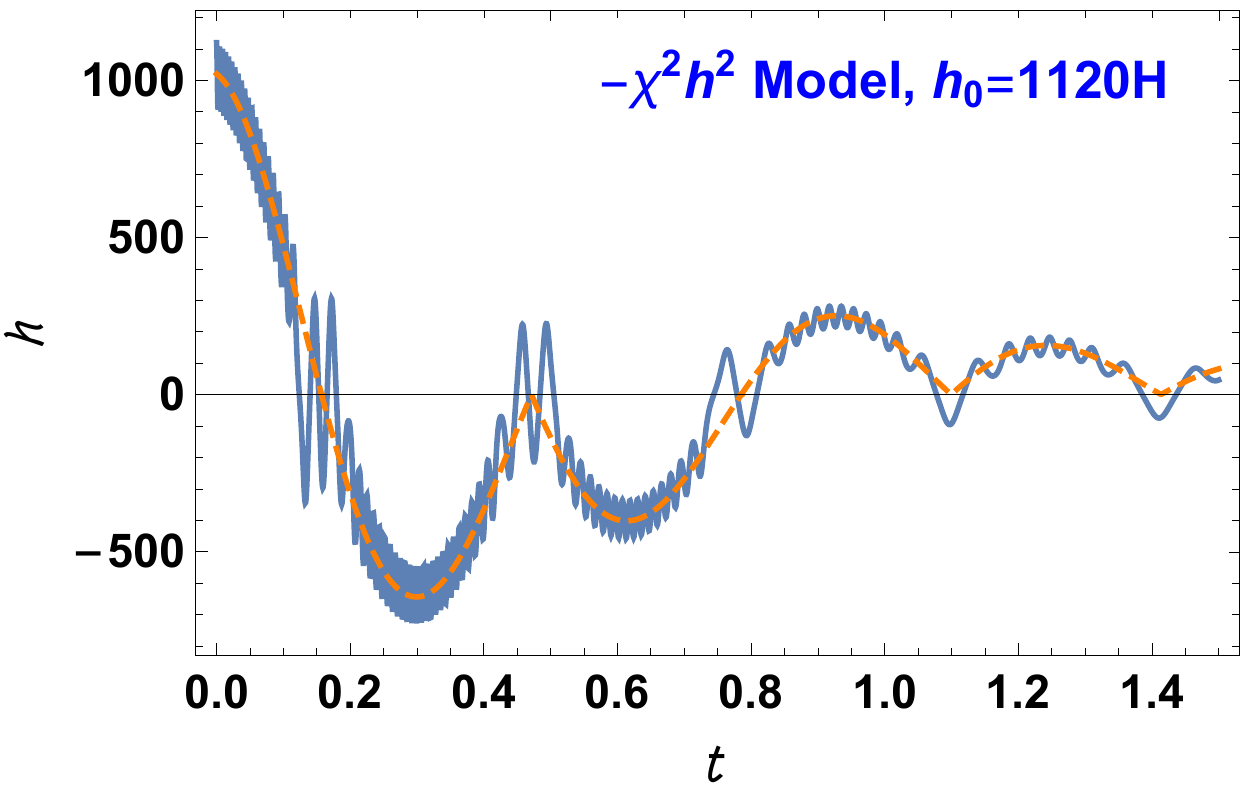}   \includegraphics[width=0.32\textwidth]{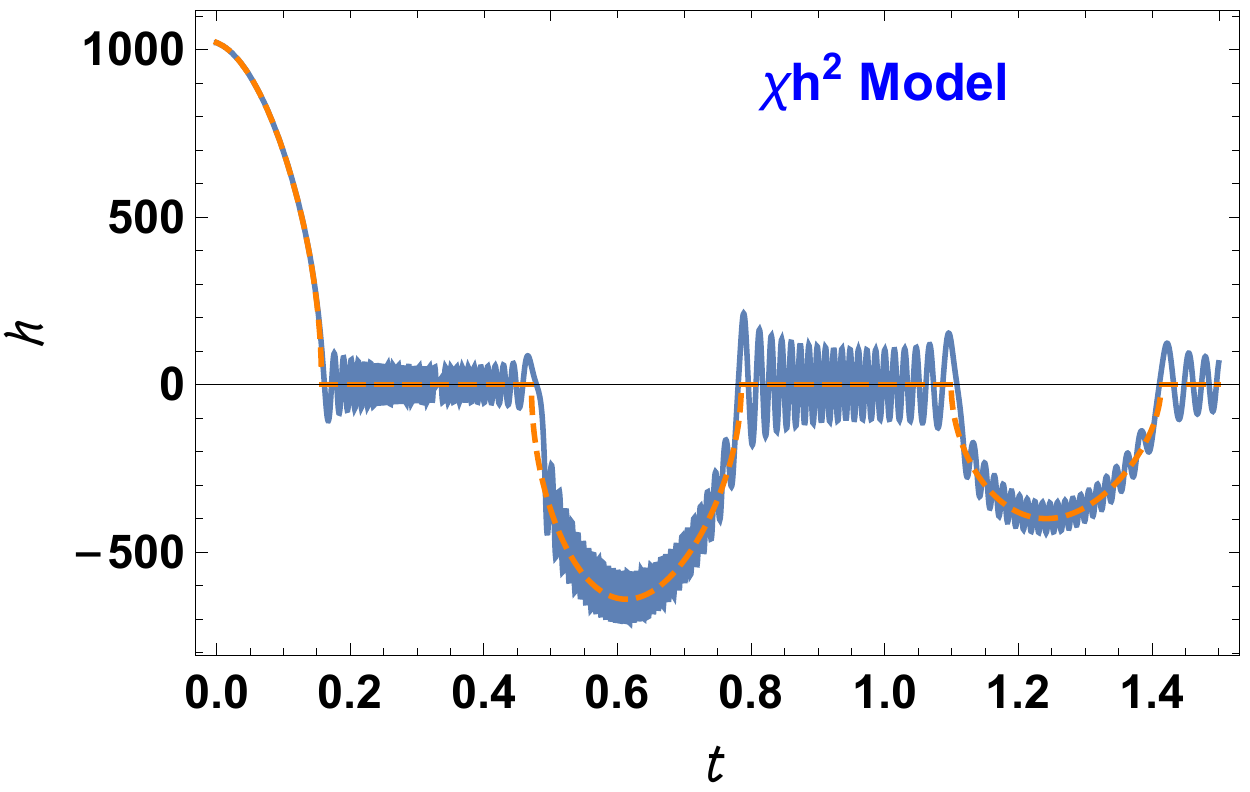}  \\
          \includegraphics[width=0.32\textwidth]{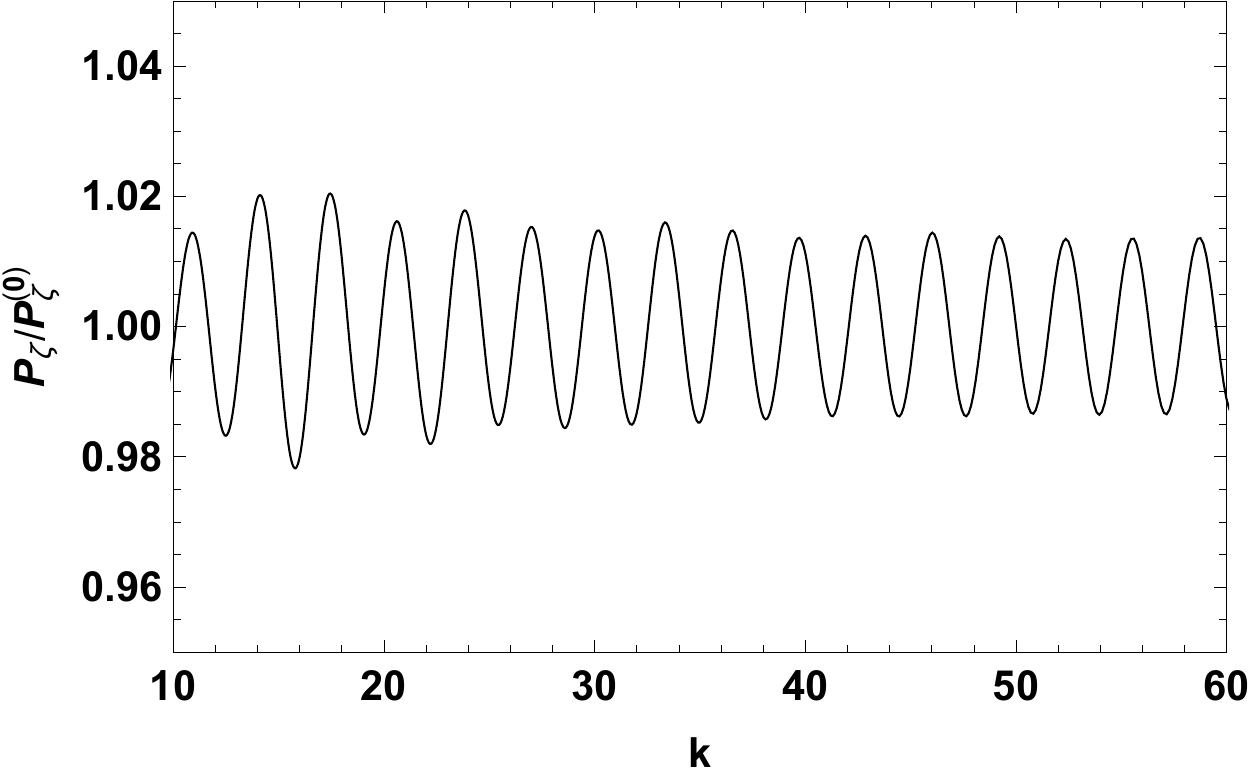} \includegraphics[width=0.32\textwidth]{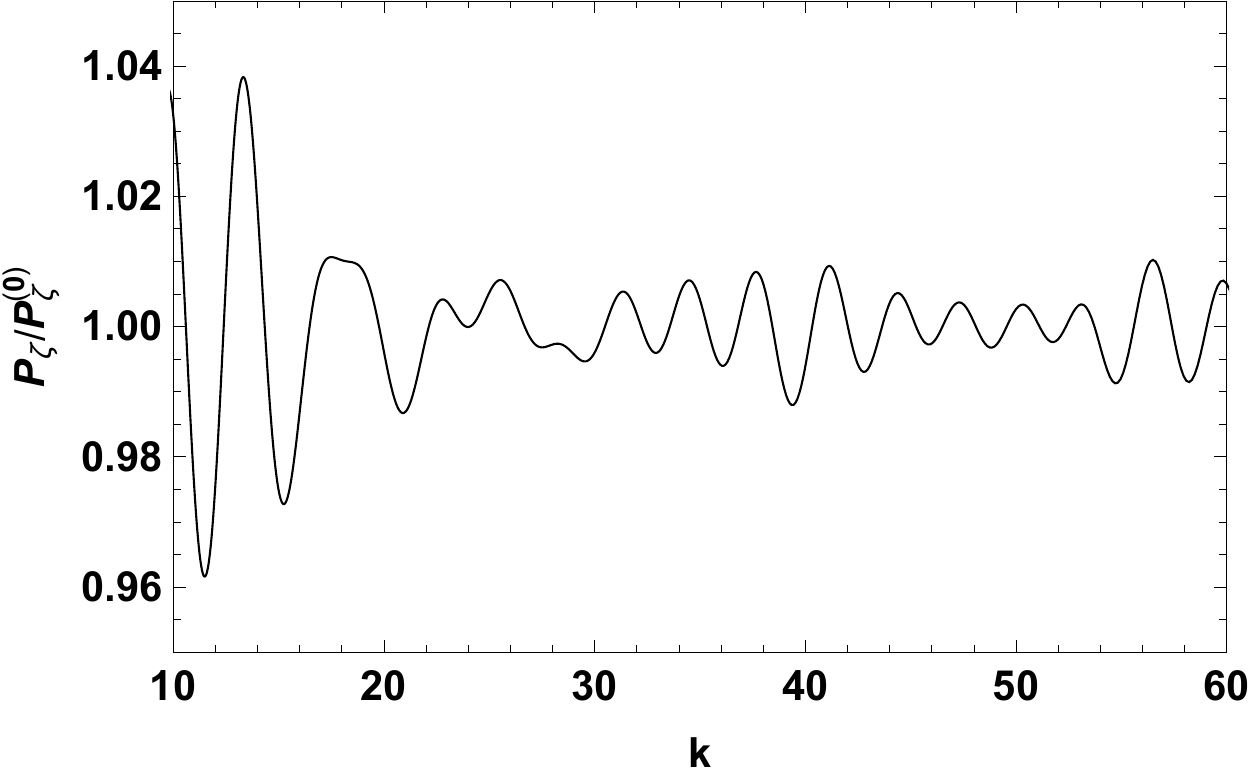}   \includegraphics[width=0.32\textwidth]{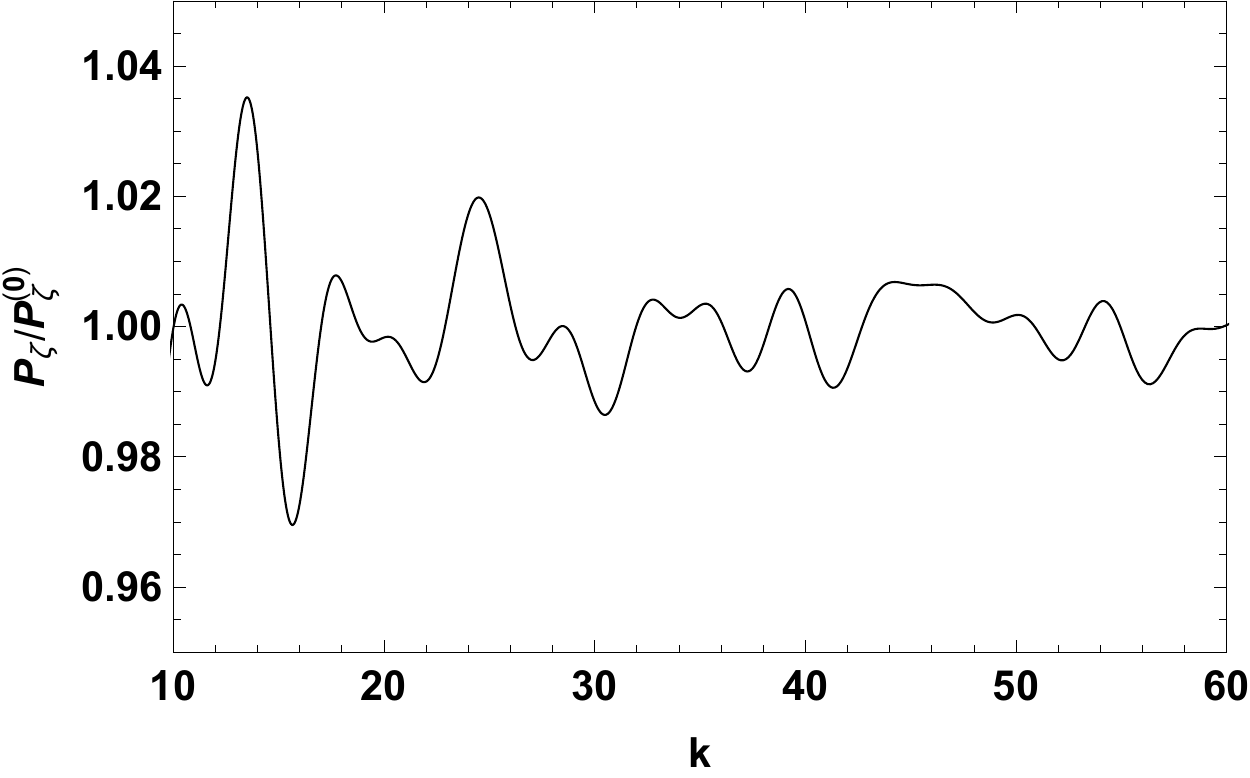}   \\
   \includegraphics[width=0.32\textwidth]{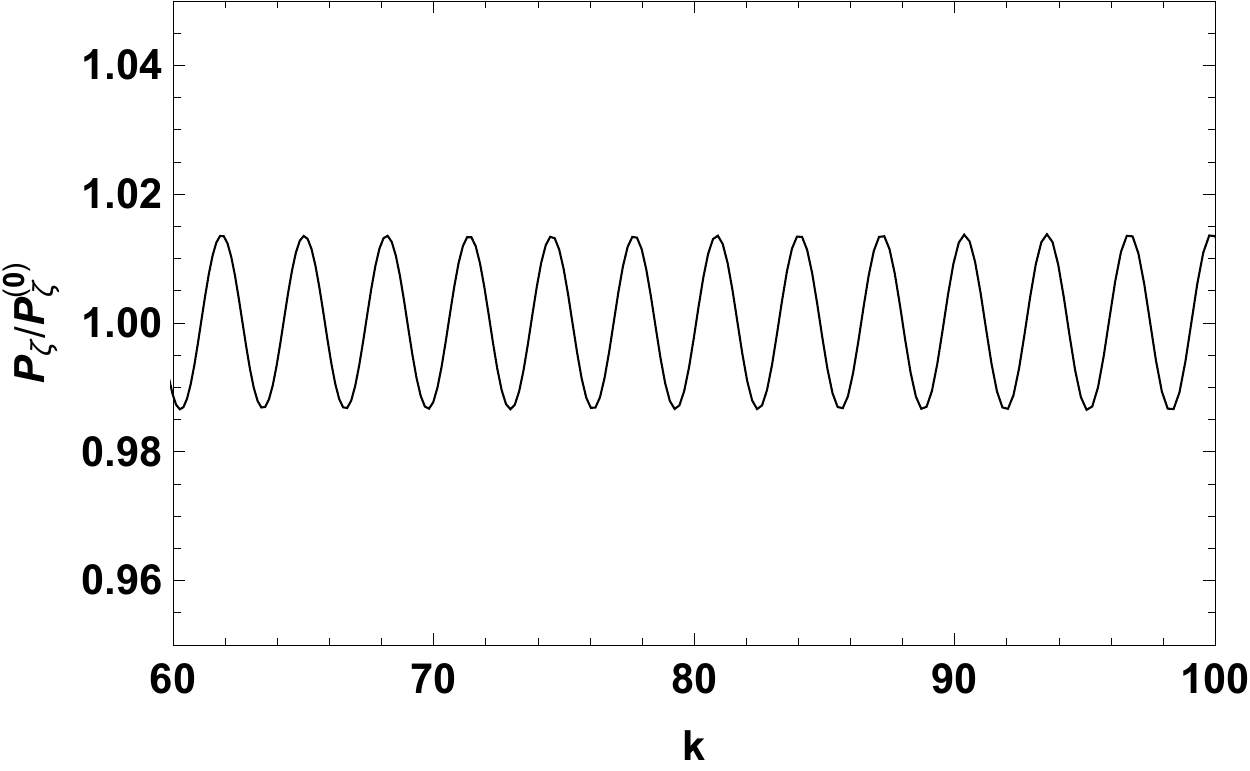} \includegraphics[width=0.32\textwidth]{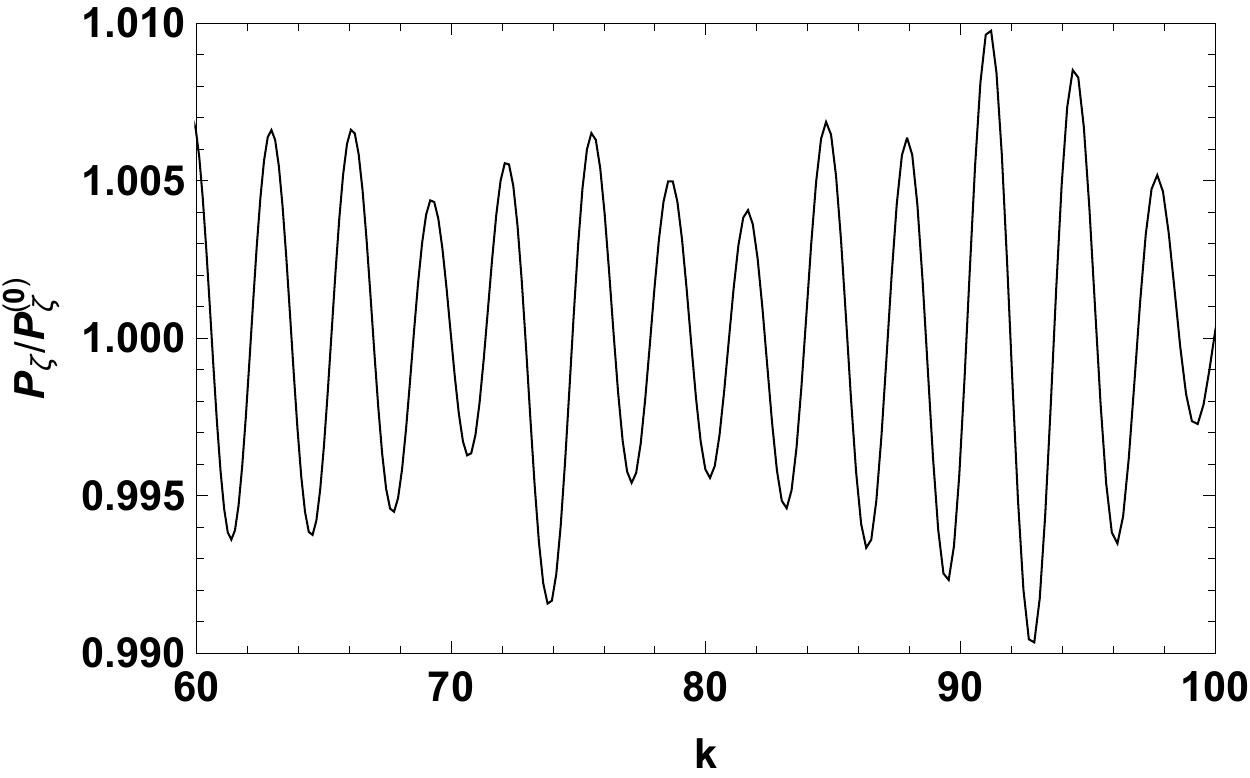}   \includegraphics[width=0.32\textwidth]{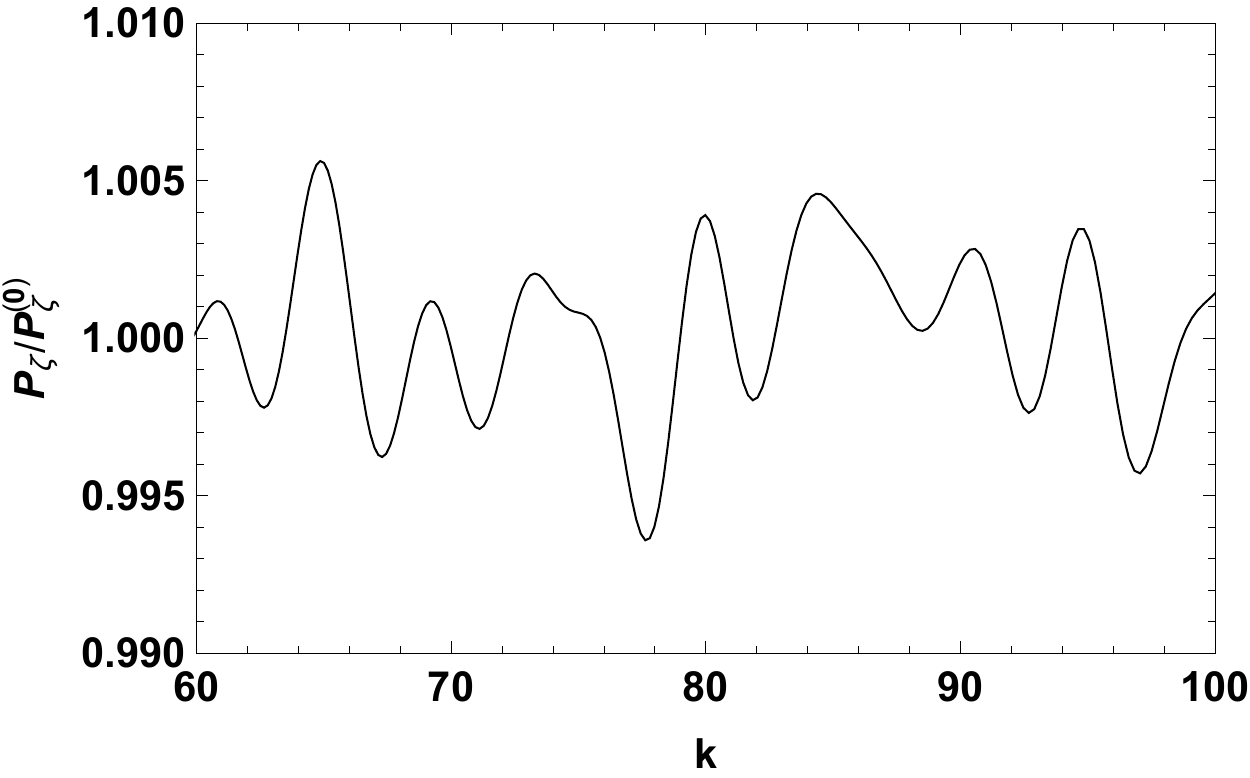}   \\                                        \includegraphics[width=0.32\textwidth]{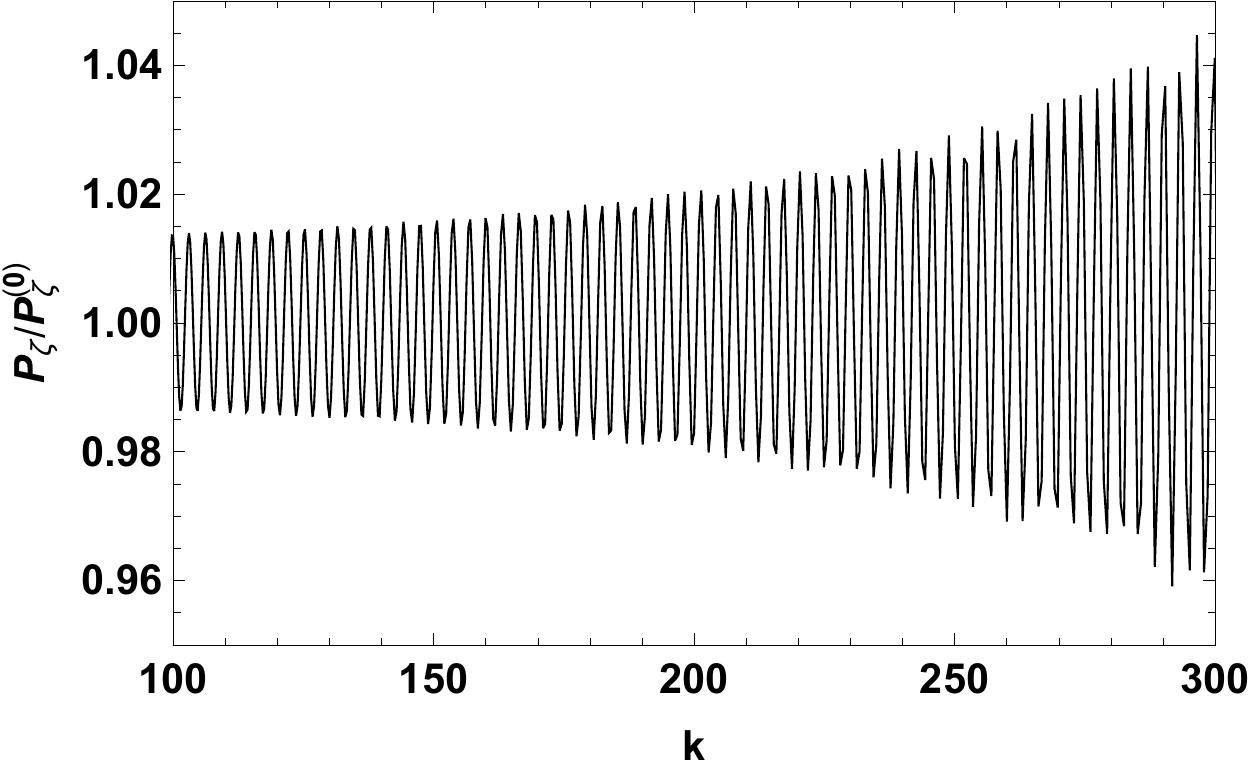} \includegraphics[width=0.32\textwidth]{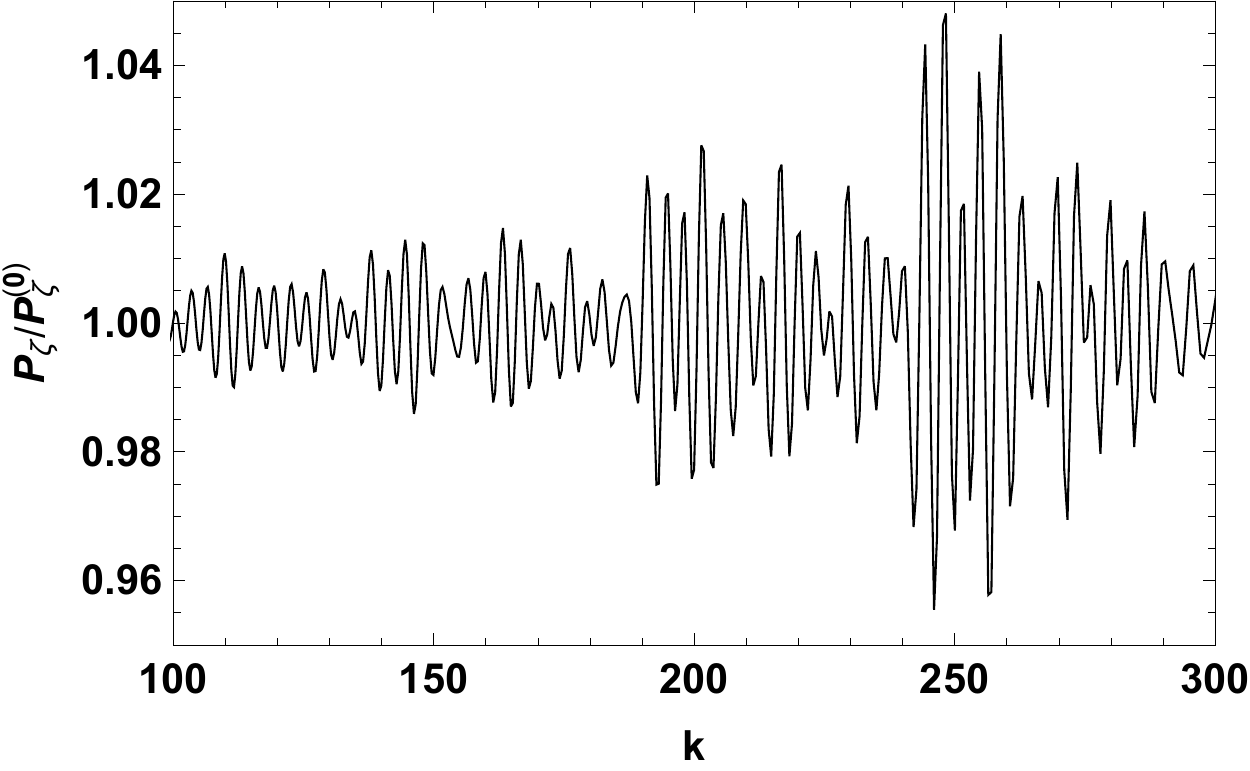} \includegraphics[width=0.32\textwidth]{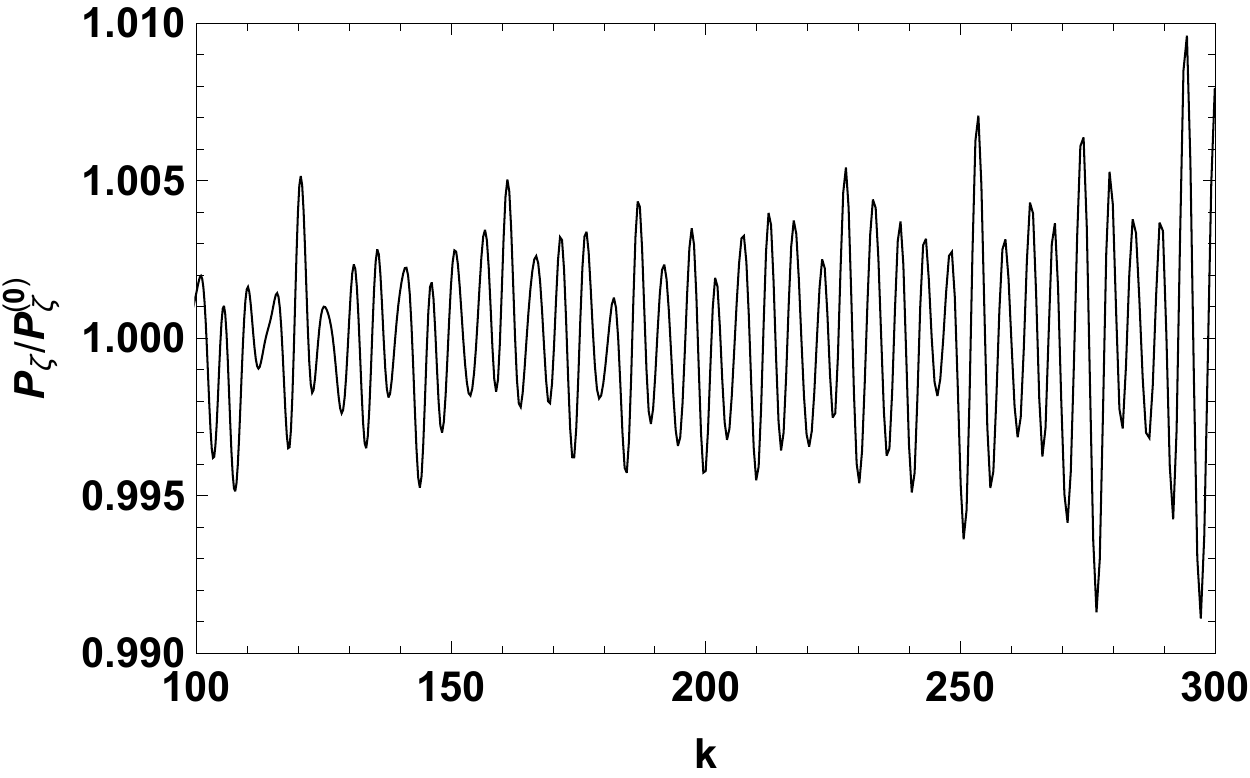}   \\
   \includegraphics[width=0.32\textwidth]{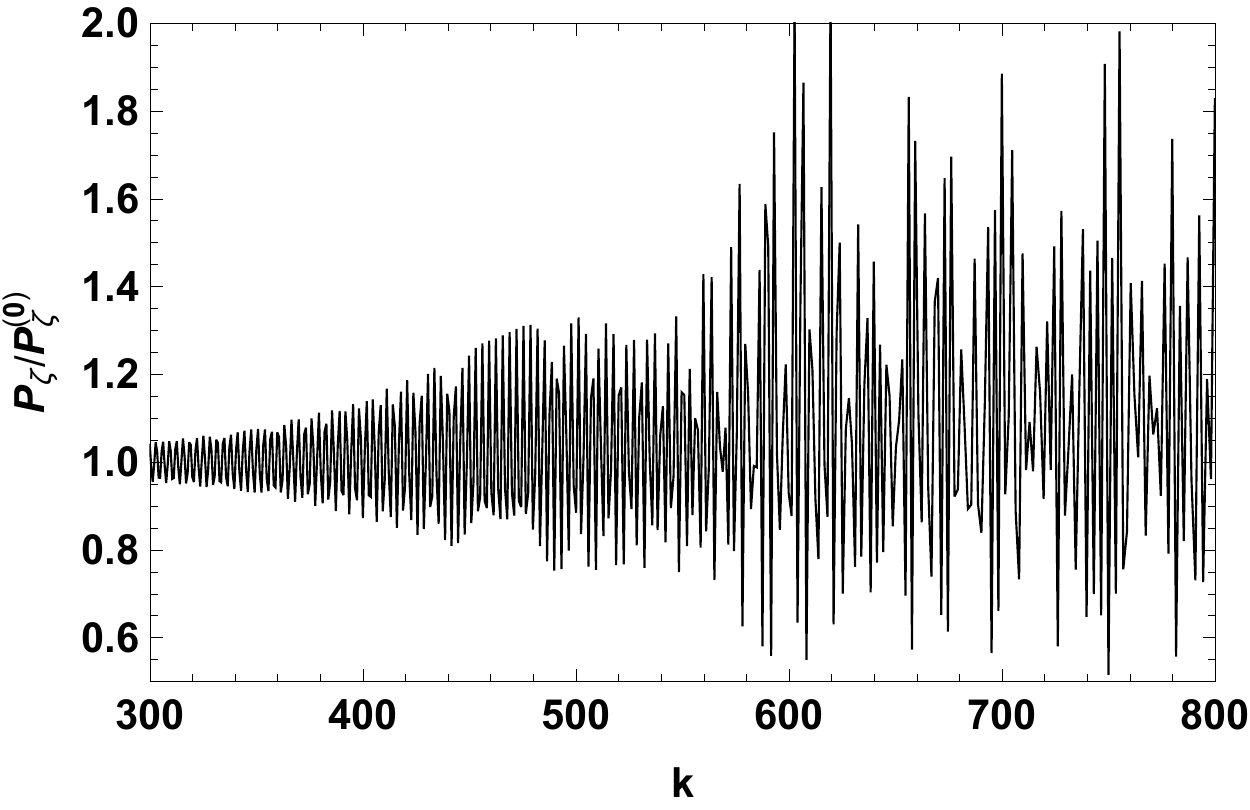} \includegraphics[width=0.32\textwidth]{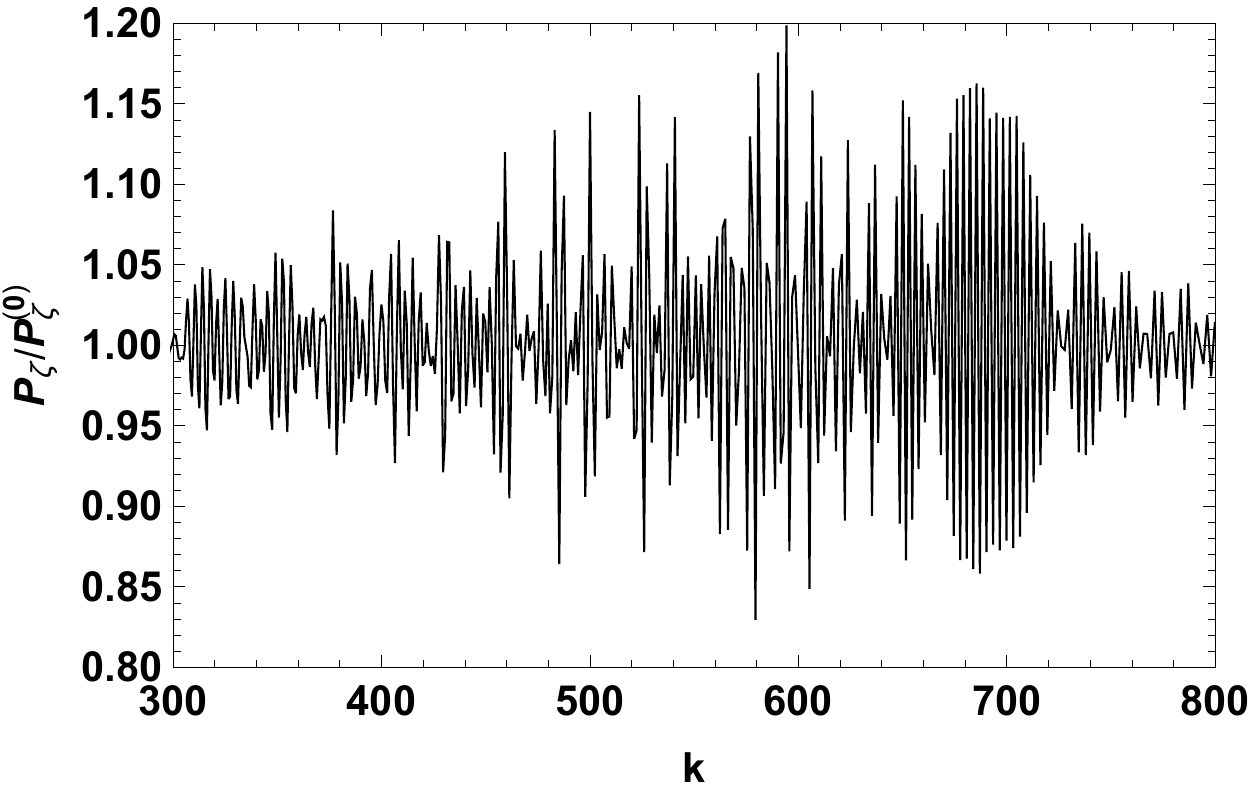}   \includegraphics[width=0.32\textwidth]{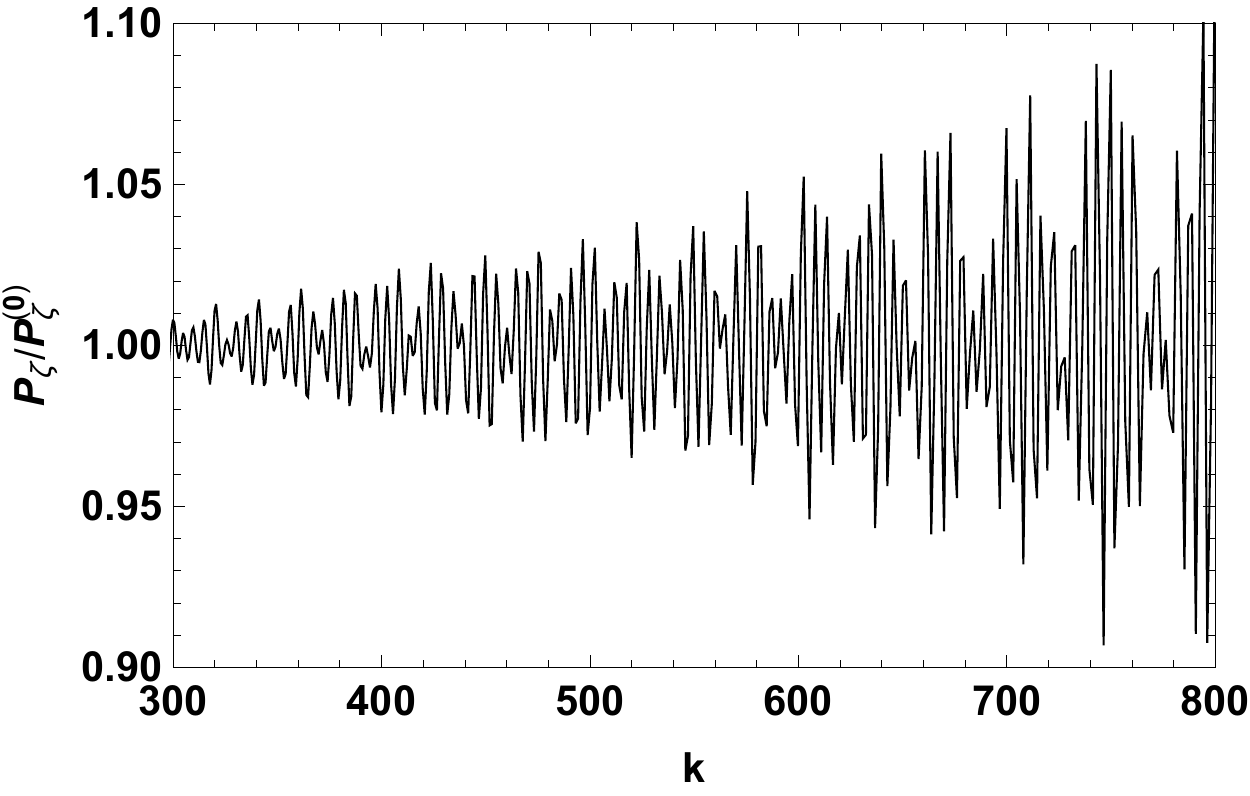}   \\     
    \caption{\label{fig:comparison} The first column correspond to $- \frac{M^2}{f^2} \chi^2 h^2$ model with the initial conditions $h_0 = 0.01 H$. The second column corresponds to $-\frac{M^2}{f^2} \chi^2 h^2$ model with $h_0 = 1120 H$; the last column corresponds to $\frac{M^2}{f} \chi h^2$. The other parameters are the same, $M = 1020H, \Lambda = 6000 H, m_\chi = 10 H$ and $\lambda = 1$. The first row is the Higgs field evolution with the orange dashed line indicating one of the instantaneous Higgs minima. The second to last rows correspond to the change in scalar two-point in the $k$ range: $(10 - 60), (60 - 100), (100 -300)$, and $(300 - 800)$. 
  \label{fig:com}  }
\end{figure}

In case {\it a)}, the Higgs field sloshes back and forth in the potential and could go from one minimum to another minimum. The duration of a Higgs oscillation is $\sim (M| \cos(m_\chi t)|)^{-1}$, so each period of the Higgs is of order $1/M$ but modulated over a longer timescale $1/m_\chi$. This is demonstrated in Fig.~\ref{fig:sloshzoomin}. In the figure, the orange curve is $M| \cos(m_\chi t)|$ while the blue curve is the Higgs value. One could see that when the Higgs potential is shallowest at $\cos(m_\chi t) \sim 0$, it is easier for the Higgs to climb over the barrier in order to go from one minimum to another and the oscillation gets fast.
These Higgs oscillations lead to a primordial spectrum which is shown in the first column of Fig.~\ref{fig:com}. From it, one could see that in the low $k$ range, the oscillations are described by a single periodic function with periodicity set by $1/H$ while at large $k$, the primordial spectrum demonstrates an irregular oscillation pattern. In the entire $k$ range, the oscillation features are very different from those in the phase oscillation model. 

In case {\it b)}, the Higgs oscillates mostly around one minimum with small oscillation amplitude except when the modulus crosses its origin and the Higgs potential is flattened. At those times, Higgs oscillates around the origin and could switch from one minimum to another. In this case, the Higgs field value could also be approximated as $h = h_{\rm vev} + h_{\rm osc}$, analogous to what happens in the broken phase of the phase oscillation model. Indeed, the resulting oscillations in the primordial spectrum, shown in the middle column of Fig.~\ref{fig:com}, have some similar features as the phase oscillation model. In the low $k$ range, the oscillations are superpositions of oscillations with slightly different frequencies while in the high $k$ range, there also exist some k-wavepackets. Yet the low $k$ spectrum ($k \lesssim 100$) originates from a different $h_{\rm vev}^2$ profile and thus is different from that of the phase oscillation model. In the high $k$ range ($k \gtrsim 400$), the wavepackets appear more irregular and chaotic. 

In case {\it b)}, we choose the Higgs to start close to but not exactly at the minimum of its potential. If Higgs starts from the vacuum initially, as we choose for the phase oscillation model, Higgs trajectory will track the true instant minimum in the symmetry breaking model and $h_{\rm osc} \approx 0$. On the other hand, in the phase oscillation model, although we choose Higgs to start from the true minimum, fast oscillations around the instant minimum with frequency $\sim M$ will be generated, as shown in the third plot in the first row of Fig.~\ref{fig:com}.

\section{Conclusion and outlook}
\label{sec:con}

To conclude, we have investigated the cosmological signatures of spontaneously broken and restored symmetries of a Higgs field during inflation. For this purpose, we have constructed a toy model, in which the oscillating modulus field controls the sign of the Higgs mass squared, and thus the oscillatory transitions between the symmetric and broken phases of the Higgs. 

Rich cosmological phenomenology arises from this model. The piecewise half-cosine oscillation of the Higgs vev and the more rapid oscillation of the Higgs field around the local minima provide distinct features at different scales on the primordial power spectrum. Some model parameters can be fit from these features. Those features may be observed in the future CMB, large scale structure and 21 cm experiments. If the features are measured very precisely, there may even be hope to probe the existence of high energy scales beyond the weak scale and whether the Higgs mass is fine-tuned or not.

There are many interesting possible directions that we do have left unaddressed in this paper, including:

\textit{Non-Gaussianities}. The resonant signal should be a natural source for large non-Gaussianities \cite{Chen:2008wn, Flauger:2014ana}. The current constraints \cite{Ade:2015ava} and future potentials (see, for example SphereX \cite{Dore:2014cca}) can provide more information about the model parameters.

\textit{Non-linear decay of the deep sub-horizon excitations}. In our model, due to the Higgs oscillation, the resonance of the curvature perturbation happens at deep subhorizon scales. For the fluctuation modes to evolve until horizon exit, the mode may decay to longer wave length modes through non-linear dynamics \cite{Jiang:2015hfa, Jiang:2016nok}. The significance of this effect depends on the interaction details of the inflaton sector and is omitted in the current study. The decay will also result in a characteristic shape of non-Gaussianity \cite{Jiang:2016nok}.

\textit{Relation to the particle physics SM}. In this paper we have studied the phenomenology of symmetry breaking and called the corresponding field the ``Higgs''. But we did not get involved in the particle physics details of whether the field is a standard model Higgs, or even if the field carries any gauge charge. Cosmological collider physics \cite{Chen:2009zp,Arkani-Hamed:2015bza} and related phenomenological studies of inflationary signals of the particle physics SM \cite{Chen:2016nrs,Chen:2016uwp,Chen:2016hrz,Kumar:2017ecc} are expected to help tell if the cosmological signature comes from the SM or not. It will be interesting to get a more complete picture taking the model building details into consideration.

\textit{Quantum probes of the Higgs potential}. One drawback of our current work is that we need some event to trigger the modulus oscillation. Though such possibilities are abundant, they are considered to be additions to the minimal model. Actually, even if the modulus does not oscillate classically, it still has quantum zero-point oscillations which may lead to smaller, but more universal effects similar to \cite{Chen:2015lza}. The effect may be more visible in high scale inflation unlike our current study of low scale inflation, because that allows the modulus field to oscillate with a higher quantum zero-point amplitude. Also, it is interesting to see if the Higgs potential and its fine-tuning problem can be probed by landscape tomography \cite{Chen:2018uul, Chen:2018brw} in a symmetry broken phase, even if the modulus oscillation amplitude does not bring us to the symmetry restoring phase transition.

\acknowledgments{We thank Yue Zhao for collaboration in the early stage of the project. We thank Xingang Chen for detailed comments on the first preprint version of this paper. We thank Manuel Buen-Abad, Mustafa Amin, Andrew Cohen, Benjamin Wallisch and Henry Tye for useful discussions. JF is supported by the DOE grant DE-SC-0010010 and NASA grant 80NSSC18K1010.  MR is supported in part by the DOE grant DE-SC0013607 and the NASA grant NNX16AI12G. YW  is supported in part by
ECS Grant 26300316 and GRF Grant 16301917 and 16304418 from the Research Grants
Council of Hong Kong.}

\appendix

\section{Higgs oscillations}
\label{app:higgs}
In this appendix, we present more details of the analytical estimate for the time evolution of the Higgs field. 
First, we could adopt the WKB approximation to study $h_{\rm osc}$ as $|m_\mathrm{eff}| \sim M \gg m_\chi$ in our parameter space when $t$ is away from the transition points. Ignoring Hubble friction and higher order terms of order ${\cal O}(h_{\rm osc}^2)$, we have, away from the transition times, 
\begin{align}
\ddot{h}_{\rm osc} + m_{\rm eff}^2(t) h_{\rm osc} = 0, \quad & \quad m_{\rm eff}^2 > 0,  \\
\ddot{h}_{\rm osc} -2 m_{\rm eff}^2(t) h_{\rm osc} = 0, \quad & \quad m_{\rm eff}^2 < 0.
\end{align}
In deriving the second equation, we also dropped the $\ddot{h}_{\rm vev}$ term, which is of order $m_\chi^2M \ll \ddot{h}_{\rm osc}$. 
The WKB solutions are then 
\begin{align}\label{eq:higgs-osc}
  &  h_\mathrm{osc} = A(t) e^{i\theta(t)}~,
    \qquad
    \theta(t) \equiv \sqrt{2} \int^t |m_\mathrm{eff}(t')| dt'~, \quad m_{\rm eff}^2 < 0, \\
   &     h_\mathrm{osc} = \tilde{A}(t) e^{i\tilde{\theta}(t)}~,
    \qquad
    \tilde{\theta}(t) \equiv  \int^t |m_\mathrm{eff}(t')| dt'~, \quad m_{\rm eff}^2 > 0
\end{align}
where $A(t), \tilde{A}(t)$ are slowly varying functions compared to the $m_\mathrm{eff}$ scale. There are also negative frequency modes. They do not contribute to the resonances that dominate the correction to the inflaton spectrum and thus we ignore them. 

To understand the oscillations near the transition times $t_{\rm s}$ when $m_{\rm eff}^2 = 0$ (moving from positive to negative values), we will linearize the equations and write 
\begin{align}
m_{\rm eff}^2 (t) =\left. \frac{d m_{\rm eff}^2}{dt} \right |_{t = t_s} (t -t_s) \approx - M^2 m_\chi e^{-3Ht_s/2} (t-t_s),
\end{align}
where we ignore $m_h^2$, thus approximating $\cos(m_\chi t_s) = 0$, and assume $m_\chi \gg H$. Then the linearized equations are 
\begin{align}
& \ddot{h}_{\rm osc} - \beta^3 (t-t_s) h_{\rm osc} = 0, \quad \quad  \quad \quad t-t_s = 0^-,  \\
& \ddot{h}_{\rm osc} +2 \beta^3 (t-t_s) h_{\rm osc} = 0, \quad  \quad \quad \; \; t-t_s = 0^+, \\
& {\rm where} \quad  \beta= (M^2 m_\chi)^{1/3} e^{-Ht_s/2}.
\end{align}
The solutions to the equations above are then
\begin{align}
h_{\rm osc} &= c_1 \Ai (\beta (t-t_s)) + c_2 \Bi(\beta (t-t_s)),  && t-t_s = 0^-, \nonumber \\
h_{\rm osc} &= \sqrt{t-t_s}\left[a_1 J_{-1/3}\left(\frac{1}{3} \left(2\beta (t-t_s)\right)^{3/2}  \right) + a_2 J_{1/3}\left(\frac{1}{3} \left(2\beta (t-t_s)\right)^{3/2}  \right) \right],   && t-t_s = 0^+, 
\end{align}
where $\Ai(x), \Bi(x)$ are Airy functions and $J_\nu(x)$ is the Bessel function of the first kind. For small arguments, these are oscillating functions with periods set by $\beta \approx (M^2 m_\chi)^{1/3}$. 

Lastly we want to develop a heuristic understanding of the amplitude of $h_{\rm osc}$, $A$ and $\tilde{A}$. The first derivative of $h=h_{\rm vev} + h_{\rm osc}$ with respect to time should be continuous through the phase transition points $t_s$. The separation into $h_{\rm vev}$ and $h_{\rm osc}$ is not well-defined near these points. In particular, the formula $h_{\rm osc} \sim \sqrt{t - t_s}$ shows that ${\dot h}_{\rm vev}$ is singular at the transition time, even though $\dot h$ is not. If we assume that the separation into $h_{\rm vev}$ and $h_{\rm osc}$ becomes meaningful at the characteristic timescale $t - t_s \sim \beta^{-1} \sim 1/(M^2 m_\chi)^{1/3}$, then a plausible assumption is that around this time the amplitudes of the two components are comparable, 
\begin{equation}
\left.h_{\rm vev} \right|_{t = t_s + \beta^{-1}} \sim \left. h_{\rm osc} \right|_{t = t_s + \beta^{-1}}.
\end{equation}
This implies that the amplitude of the oscillating component is
\begin{align}
A &\sim \left.h_{\rm vev} \right|_{t = t_s + \beta^{-1}} \sim \frac{M}{\sqrt{\lambda}} \sqrt{m_\chi \beta^{-1}} \nonumber \\
&\sim \frac{1}{\sqrt{\lambda}} \left(M^2 m_\chi \right)^{1/3}.
\label{eq:aestimate}
\end{align}
Checking a few benchmark points, we find that this analytical estimate agrees reasonably well with the numerical results. 

\section{Saddle point approximation}
\label{app:saddle}
In this appendix, we present the main formula of the saddle point approximation, which is our key tool to compute the correction to the inflaton spectrum. For an integration involving a fast oscillating function $e^{ig(t)}$ and a slowly varying function $f(t)$, the saddle point approximation gives
\begin{align}
\int_a^b dt \, e^{i  g(t)} f(t) \approx e^{i  g(t_0) + i \frac{\pi}{4} \sign g''(t_0)} f(t_0) \sqrt{\frac{2\pi}{ |g^{\prime\prime}(t_0)|}},
\end{align}
where $g^\prime(t_0) =0$. 
In our computation, the fast oscillations usually take the form $e^{i \omega t(\eta)-iq \eta}$. Then using the general formula, we have
\begin{align}
    \int_{-\infty}^0 d\eta\, e^{i\omega t(\eta) - i q \eta} f(\eta)
    \simeq
    e^{\frac{i\omega}{H} } \sqrt{2\pi i}
    \sqrt{\frac{\omega}{H q^2} }
    \left( \frac{q}{a_0 \omega}  \right)^{ \frac{i\omega}{H} }
    f \left( -\frac{\omega}{Hq} \right)~,
\end{align}
where $a_0$, the initial scale factor, is defined such that $a(t) =a_0 e^{Ht}$ and $\sqrt{i} =  e^{i\pi/4}$. 

\section{More details on $h^2_{\rm vev}$'s contribution to the inflaton wavefunction}
\label{app:vev}
In this appendix, we present more details of the computation in Sec.~\ref{sec:hvev}. Plugging Eq.~\eqref{eq:Fourier} into the last equality of Eq.~\eqref{eq:approxlatetime} and integrating by parts, we find that
\begin{align}
(k \tau)u_{k;\rm{vev}}^{(1)} &= \frac{y}{2 \Lambda^2 (2k)^{3/2}}  \frac{M^2}{\lambda}\int_{-\infty}^0 d\eta \, f(\eta) \, e^{-2ik \eta} \nonumber \\
& \int d \omega  \frac{m_\chi e^{2\pi i \omega t}}{m_\chi^2-4\pi^2\omega^2}  \left[  i  \frac{m_\chi}{2\pi \omega}+ e^{-\frac{i \pi^2 \omega}{m_\chi}} + \sum_{n} e^{-\frac{3(n+1)\pi H}{m_\chi}}\left(e^{-\frac{(3+4n) i \pi^2 \omega}{m_\chi}} +e^{-\frac{(5+4n) i \pi^2 \omega}{m_\chi}}  \right)\right], \nonumber \\
{\rm where}\quad f(\eta)&= \partial_{\eta}^2 \left[\left(1- \frac{i}{k \eta}\right)^2e^{-2ik\eta}\right] e^{2ik\eta} \quad\text{does not have rapid oscillations}.
\label{eq:appvev1}
\end{align}
Here the complementary solution at the last line of Eq.~\eqref{eq:approxlatetime} is dropped since it does not have resonant contributions. 
Then we change the order of integration: we will first integrate over $\eta$ and then $\omega$ using saddle point approximations twice. First for the $\eta$ integration, we have 
\begin{align}
\int d\eta \; e^{2i(\pi  \omega t(\eta) -  k \eta)} f(\eta)
\end{align}
By the saddle point approximation, resonance happens at 
\begin{align}
\eta_* = -\frac{\pi \omega}{H k}.
\end{align}
After integrating over $\eta$, we end up with an integration over $\omega$:
\begin{align}
&\frac{ye^{i\pi/4}}{2\Lambda^2(2k)^{3/2}}\frac{M^2m_\chi}{\lambda}\frac{\pi}{k} \int d\omega \frac{f\left(-\frac{\pi \omega}{kH}\right)\sqrt{\omega/H}}{m_\chi^2-4\pi^2\omega^2}  e^{2\pi i \frac{\omega}{H}\left(1-\ln\left(\frac{\pi \omega}{k}\right) \right)} \nonumber \\
&\times \left[  \left( i  \frac{m_\chi}{2\pi\omega}+ e^{-\frac{i \pi^2 \omega}{m_\chi}}\right) + \sum_{n} e^{-\frac{3(n+1)\pi H}{m_\chi}}\left(e^{-\frac{(3+4n) i \pi^2 \omega}{m_\chi}} +e^{-\frac{(5+4n) i \pi^2 \omega}{m_\chi}}  \right)\right].  \label{eq:appComegaintegral}
\end{align}
Notice that the prefactor in the integrand has a pole at $\omega = m_\chi/(2\pi)$. However, the integrand itself does not have a pole, because the first term in parentheses, $ i  \frac{m_\chi}{2\pi\omega}+ e^{-\frac{i \pi^2 \omega}{m_\chi}}$, and all the $n$-dependent terms in parentheses vanish at this value of $\omega$. However, if we carry out a saddle point approximation term by term, we will find resonances at the values 
\begin{align}
\text{initial approach: }\quad&\omega_* = \frac{k}{\pi}, \frac{k}{\pi} e^{-\frac{\pi H}{2m_\chi}}, \frac{k}{\pi}e^{-\frac{(3+4n)\pi H}{2m_\chi}}, \frac{k}{\pi}e^{-\frac{(5+4n)\pi H}{2m_\chi}},
\label{eq:lowkres1}
\end{align}
and the individual terms arising from these resonances will diverge at the value of $k$ corresponding to $\omega = m_\chi/(2\pi)$.

A better approach is to exploit the fact that the factors $\exp(\pm i \pi^2 \omega/m_\chi)$ vary more slowly than the factor $\exp[2\pi i \frac{\omega}{H}(1-\ln(\frac{\pi \omega}{k}))]$ or  the factor $\exp(-4ni \pi^2 \omega/m_\chi)$ (when $n$ is large). Thus, we can factor out terms that vanish when $\omega = m_\chi/(2\pi)$ (and hence cancel the pole in the denominator) and treat them  as part of the slowly varying factor in the integrand, rather than the rapidly oscillating part. In this approximation, we have resonances at the values
\begin{align}
\text{refined approach: }\quad&\omega_* = \frac{k}{\pi}, \frac{k}{\pi}e^{-\frac{2(n+1)\pi H}{m_\chi}},
\label{eq:lowkres}
\end{align}
one for each of the terms in parentheses in \eqref{eq:appComegaintegral}. The corresponding estimate for the perturbed wavefunction is
\begin{align}
(k\tau)u_{k; {\rm vev}}^{(1)} 
&\approx \frac{y}{2\Lambda^2(2k)^{3/2}}\frac{M^2m_\chi}{\lambda}  \vast\{\frac{f\left(-\frac{1}{H}\right)}{m_\chi^2-4k^2}e^{i\frac{2k}{H}}\left(\frac{i m_\chi}{2k} + e^{-i\frac{\pi k}{m_\chi}}\right) \nonumber \\
&\qquad +\sum_n 2e^{-\frac{5(n+1)\pi H}{m_\chi}}\frac{f\left(-\frac{1}{H}e^{-\frac{2(n+1)\pi H}{m_\chi}}\right)}{m_\chi^2-4k^2e^{-\frac{4(n+1)\pi H}{m_\chi}}}e^{i\frac{2k}{H}e^{-\frac{2(n+1)\pi H}{m_\chi}}}\cos\left(\frac{\pi k e^{-\frac{2(n+1)\pi H}{m_\chi}}}{m_\chi}\right)\vast\}.
 \end{align}
 Within this expression, we can further approximate $f(x) \approx -4k^2$ when $x$ is close to $-1/H$.  

\section{Sensitivity of the results to parameters} 
\label{app:paramsensitivity}

\begin{figure}[h] \centering
    \includegraphics[width=0.5\textwidth]{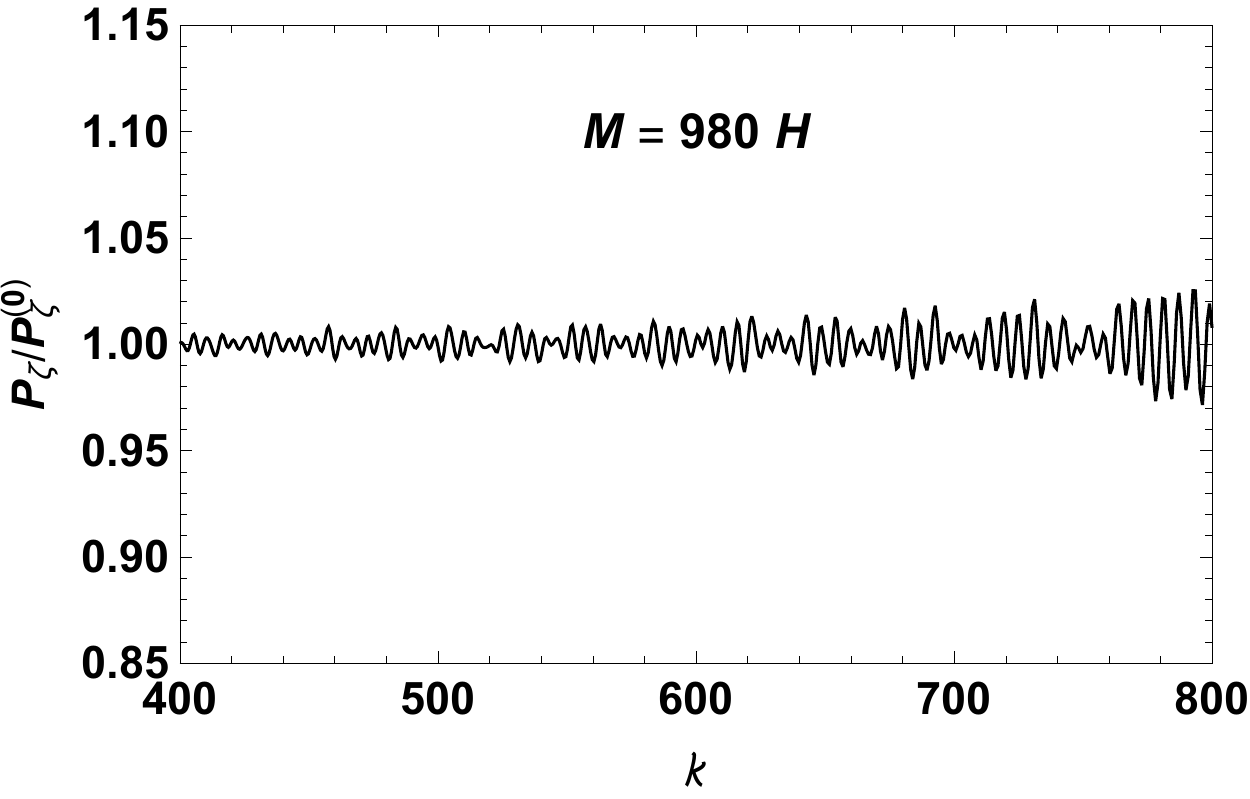}\includegraphics[width=0.5\textwidth]{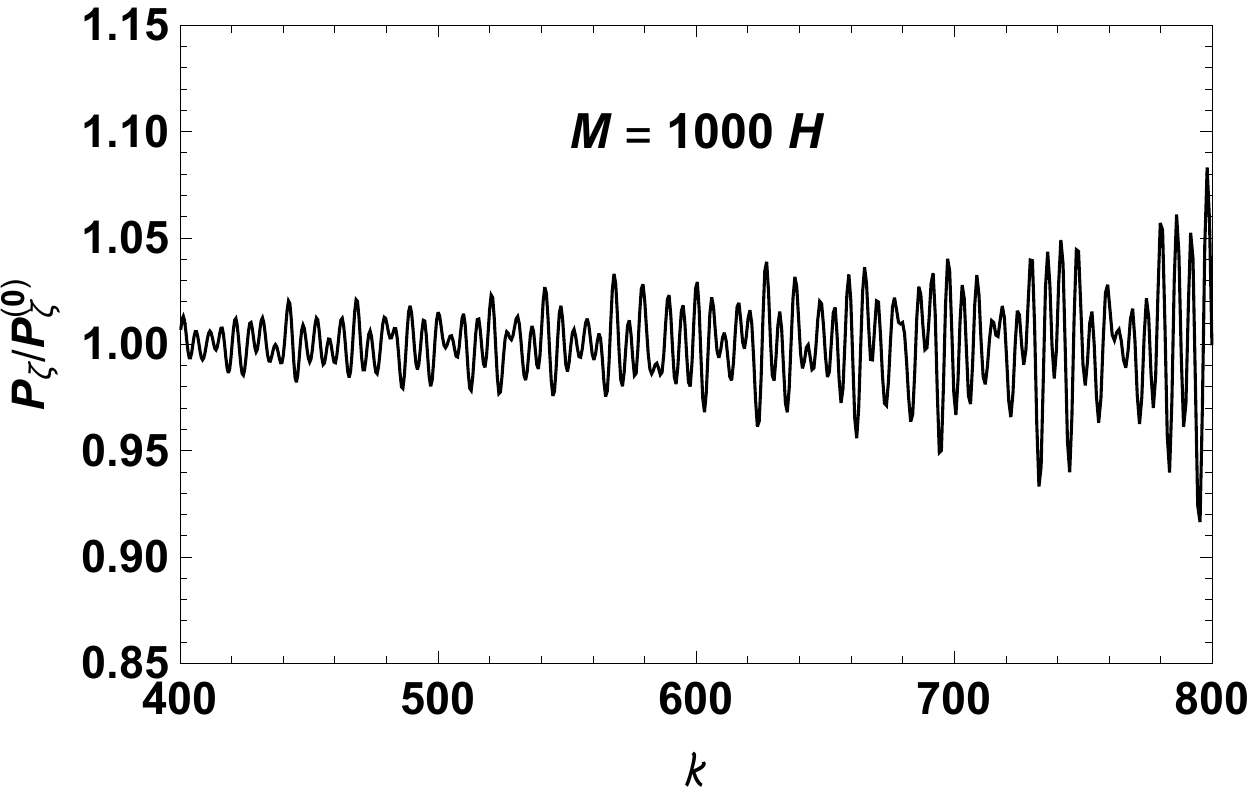}\\ 
    \includegraphics[width=0.5\textwidth]{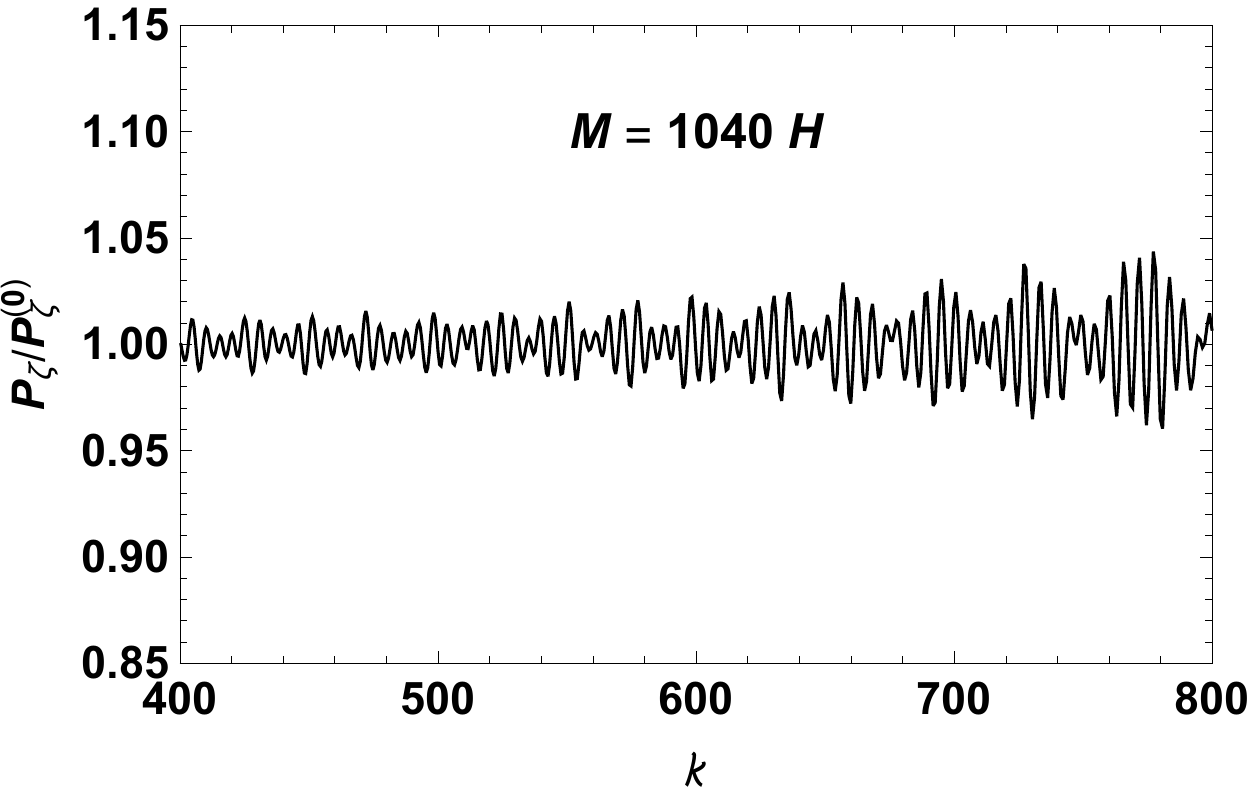}\includegraphics[width=0.5\textwidth]{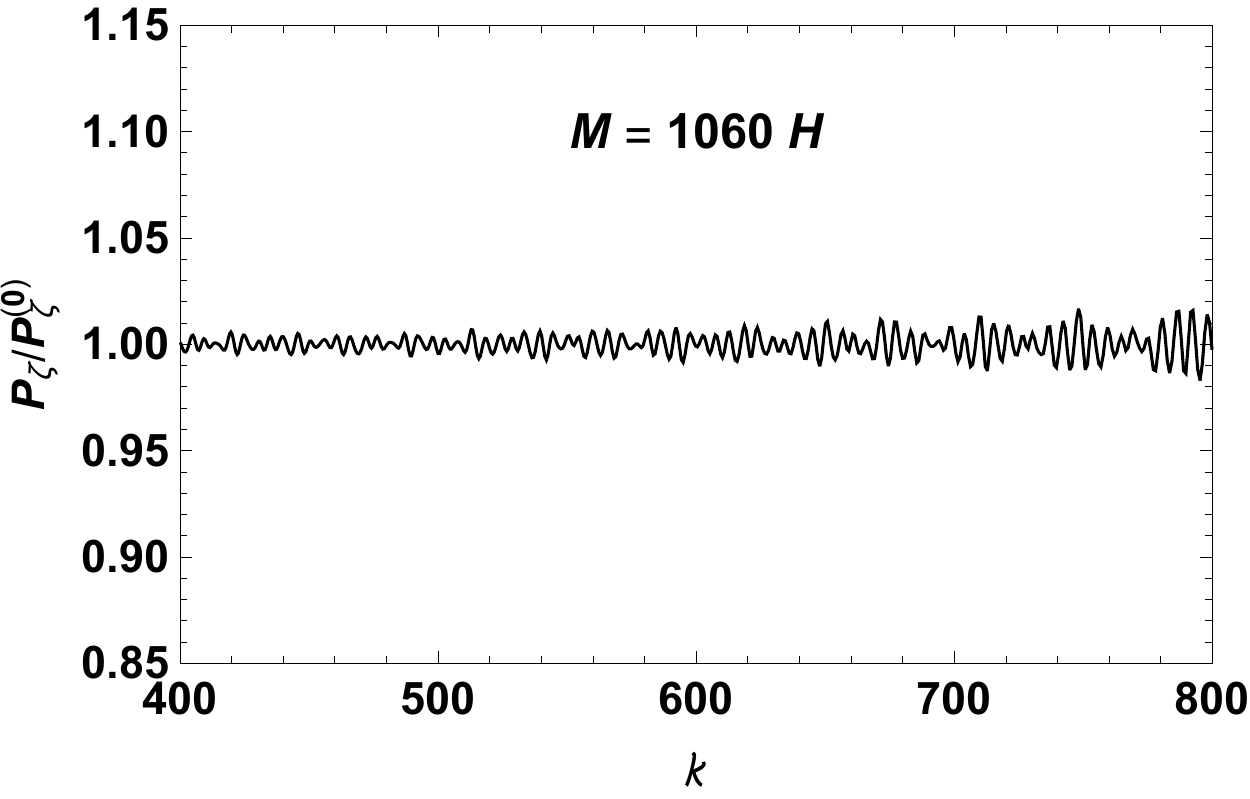}
    \caption{\label{fig:highkvariations} Like Fig.~\ref{fig:highk}, this shows the ratio $P_\zeta/P_\zeta^{(0)}$ in the high-$k$ range from solving the system numerically. However, now we show the result for a few different choices of $M$ (keeping the axes fixed in each plot): the top two plots have smaller values $M = 980 H, 1000H$ and the lower two plots have larger values $M = 1040 H, 1060 H$.
    }
\end{figure}

We remarked in \S\ref{sec:summaryofsignatures} that the details of the high-$k$ spectrum are sensitive to the choice of parameters. We illustrate this in Fig.~\ref{fig:highkvariations}, which shows four panels that are computed in precisely the same way as our earlier Fig.~\ref{fig:highk} except for the choice of $M$. We see that none of these four neighboring choices of $M$ produce as large an amplitude as we saw in Fig.~\ref{fig:highk} with the choice $M = 1020 H$. Furthermore, in the case $M = 1000 H$, which has nearly as large an amplitude, we see that the k-wavepacket features are less distinct; they appear more ``noisy'' or chaotic. The other cases all show sharply-defined wavepackets, but with smaller amplitude. Qualitatively, the properties are all as in the benchmark that we discussed in the main text, and the underlying reason for the features remains as discussed in \S\ref{sec:hvevhosc}. The variations seem to be due to the fact that the relative size of $h_{\rm vev}$ and $h_{\rm osc}$ in a given cycle depends on the precise timing of when, in a given cycle of $h_{\rm osc}$, the phase transition occurs. This timing affects the initial velocity, and hence the amplitude, with which the Higgs oscillates around $h_{\rm vev}$ in the next cycle. Because a relatively small change in $M$ can produce a noticeable change in the spectrum at large $k$, accurately inferring the underlying parameters from data may be complicated. 

\section{Comparison with the model in which Higgs is always in the symmetric phase}
\label{app:compare}
In this appendix, we compare the phase oscillation model with another class of model in which the Higgs always stays in the symmetric phase (in contrast to \S\ref{sec:compare}, where we assumed that the Higgs was always in the broken phase). We still assume that Higgs couples to the inflaton through the $h^2(\partial \phi)^2$ operator. The simplest possibility in this class is that the Higgs has a constant positive mass term and no self-coupling $V(h) = \frac{1}{2} M^2 h^2$. In this case, the Higgs value is a simple cosine function
\begin{align}
h^2(t) = h_0^2 e^{-3Ht} \cos^2(M t) = \frac{h_0^2}{2} e^{-3Ht} \left(1+ \cos(2Mt)\right),
\end{align}
where $h_0$ is the initial value of the Higgs field. Using the saddle point approximation, we find that resonance happens at 
\begin{align}
\frac{k}{a(t)} = M,
\end{align} 
and for $k> M$, 
\begin{align}
\frac{P_\zeta(k)}{P_\zeta^{(0)}(k)} -1\propto  k^{-3} \sin \left[\frac{2M}{H}\ln \left(\frac{k}{M }\right)+{\rm constant \, phase}\right]. 
\end{align}
The $\sin (C \log k)$ pattern has been known in the literature to be the standard clock signal for a heavy oscillating field coupling to the inflaton~\cite{Chen:2011zf}. 

\begin{figure}[h] 
\begin{center}
 \includegraphics[width=0.32\textwidth]{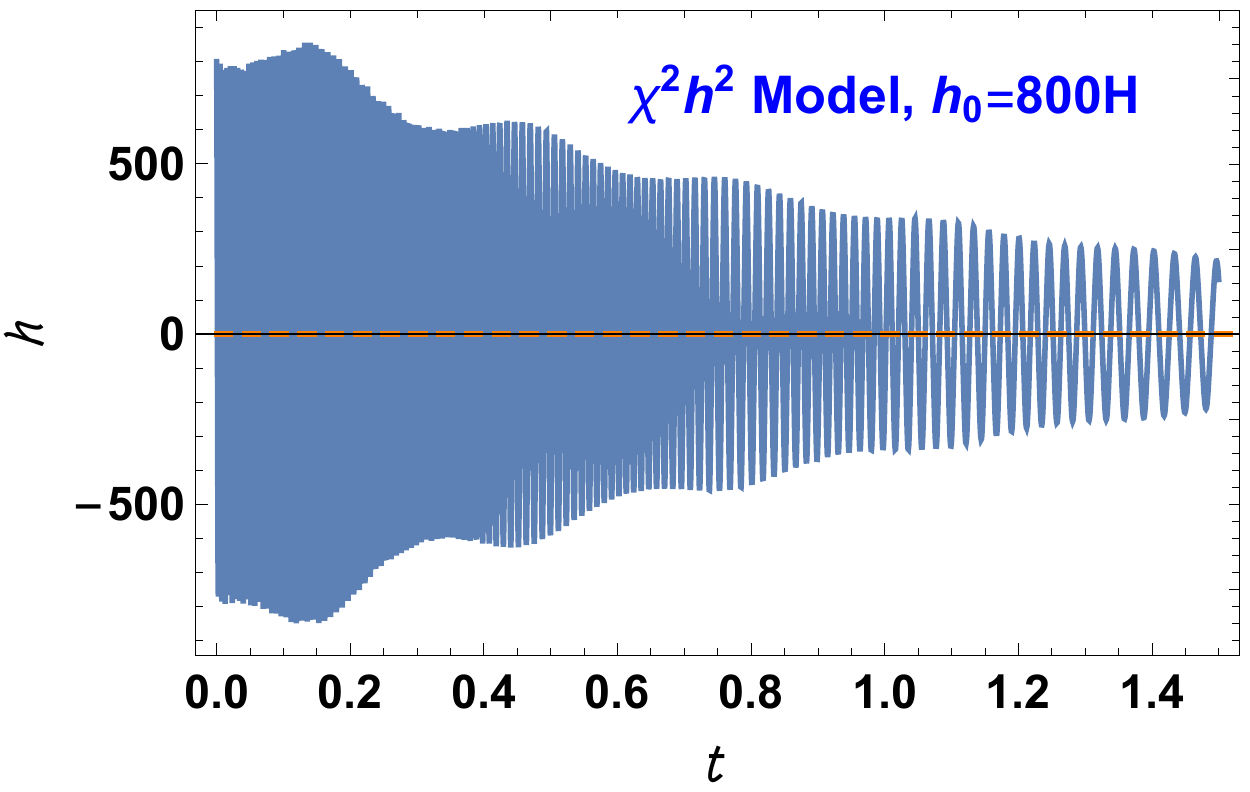}   \includegraphics[width=0.32\textwidth]{hevolmodel1cond1}  \\
\includegraphics[width=0.32\textwidth]{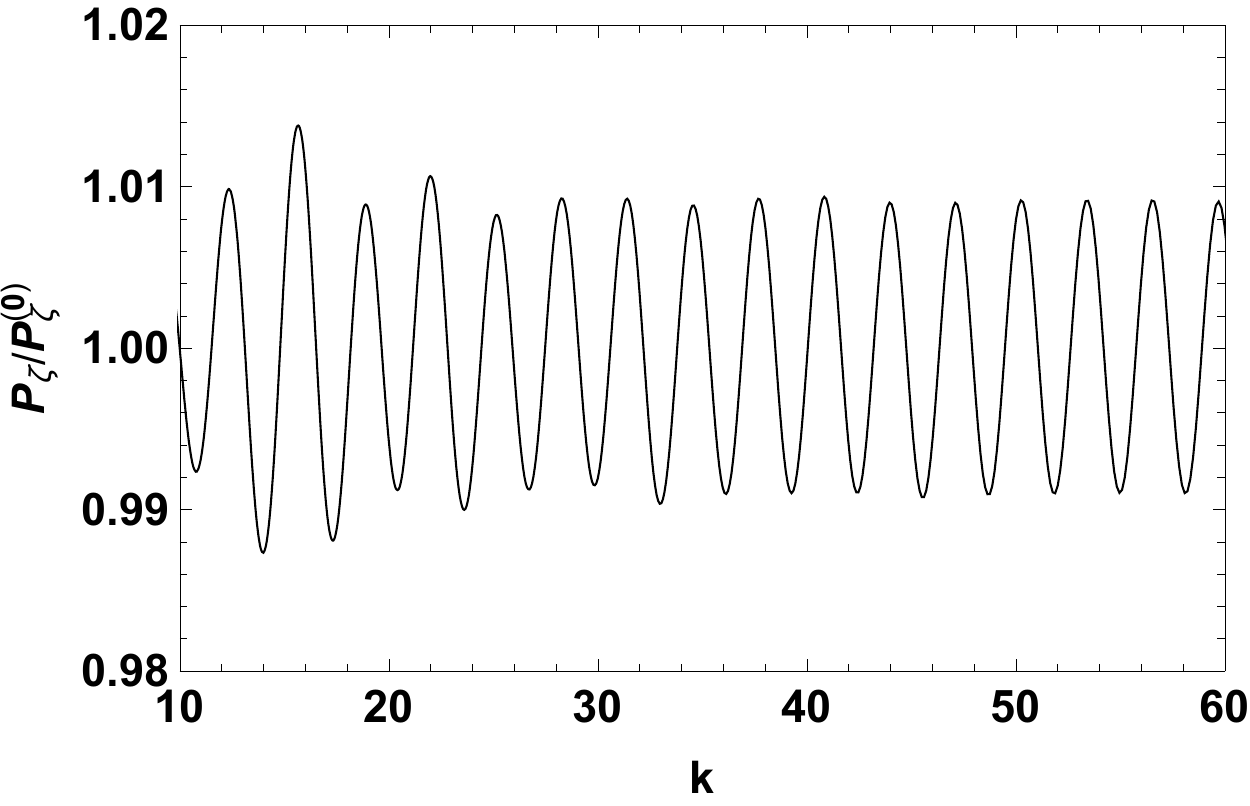}   \includegraphics[width=0.32\textwidth]{model1plot2}   \\
 \includegraphics[width=0.32\textwidth]{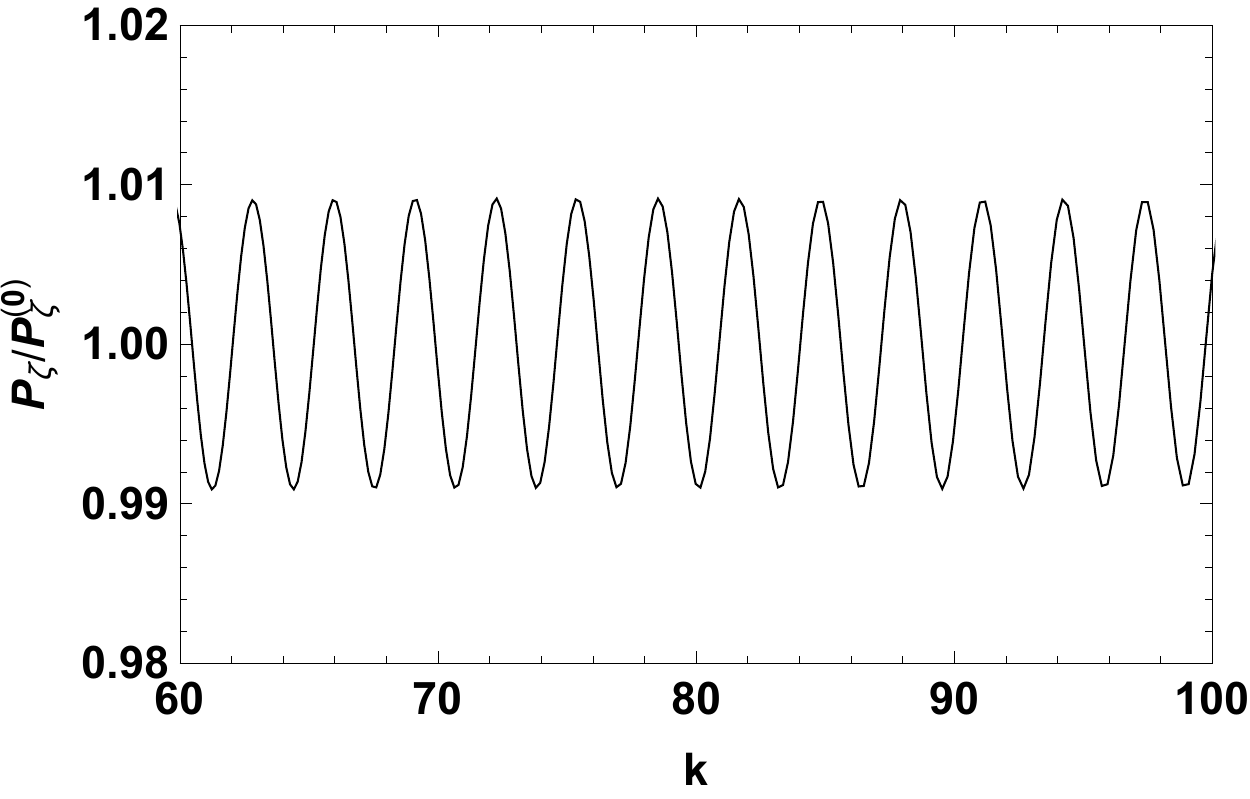}   \includegraphics[width=0.32\textwidth]{model1plot3}   \\                                      
 \includegraphics[width=0.32\textwidth]{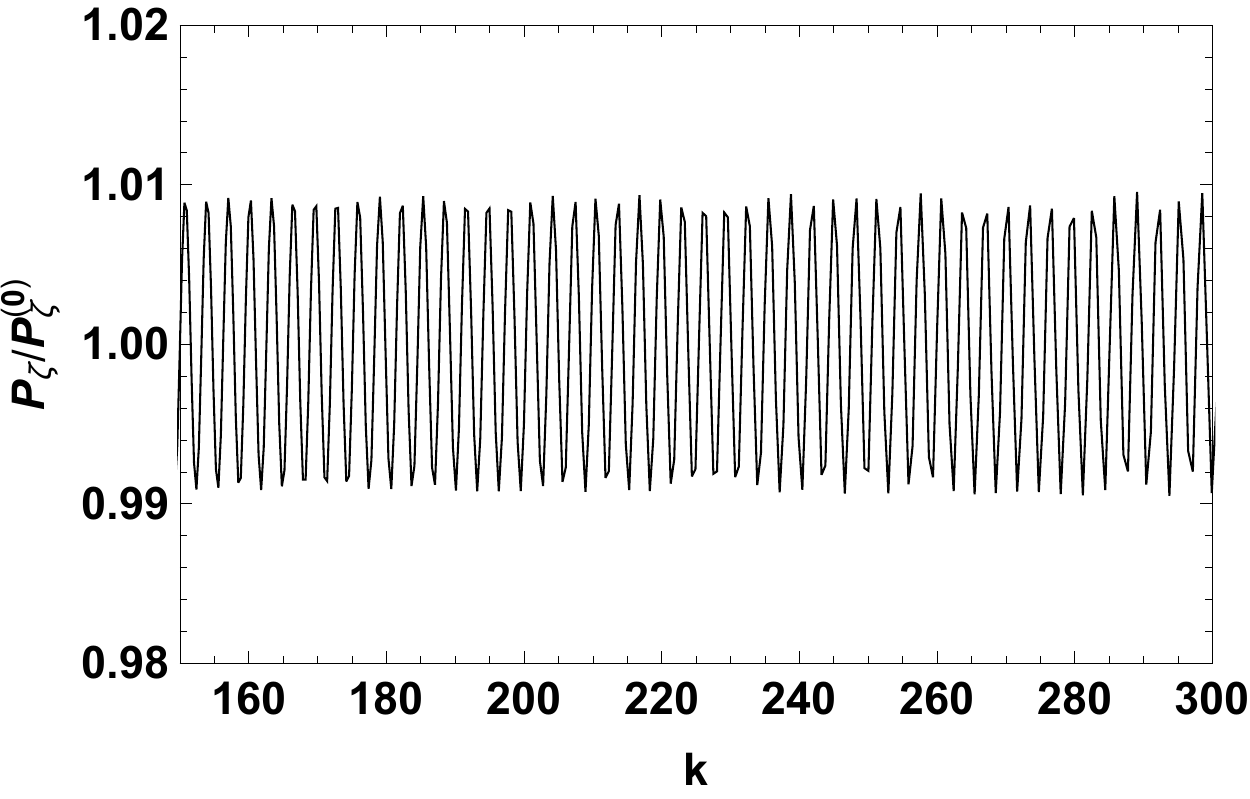} \includegraphics[width=0.32\textwidth]{model1plot4}   \\
\includegraphics[width=0.32\textwidth]{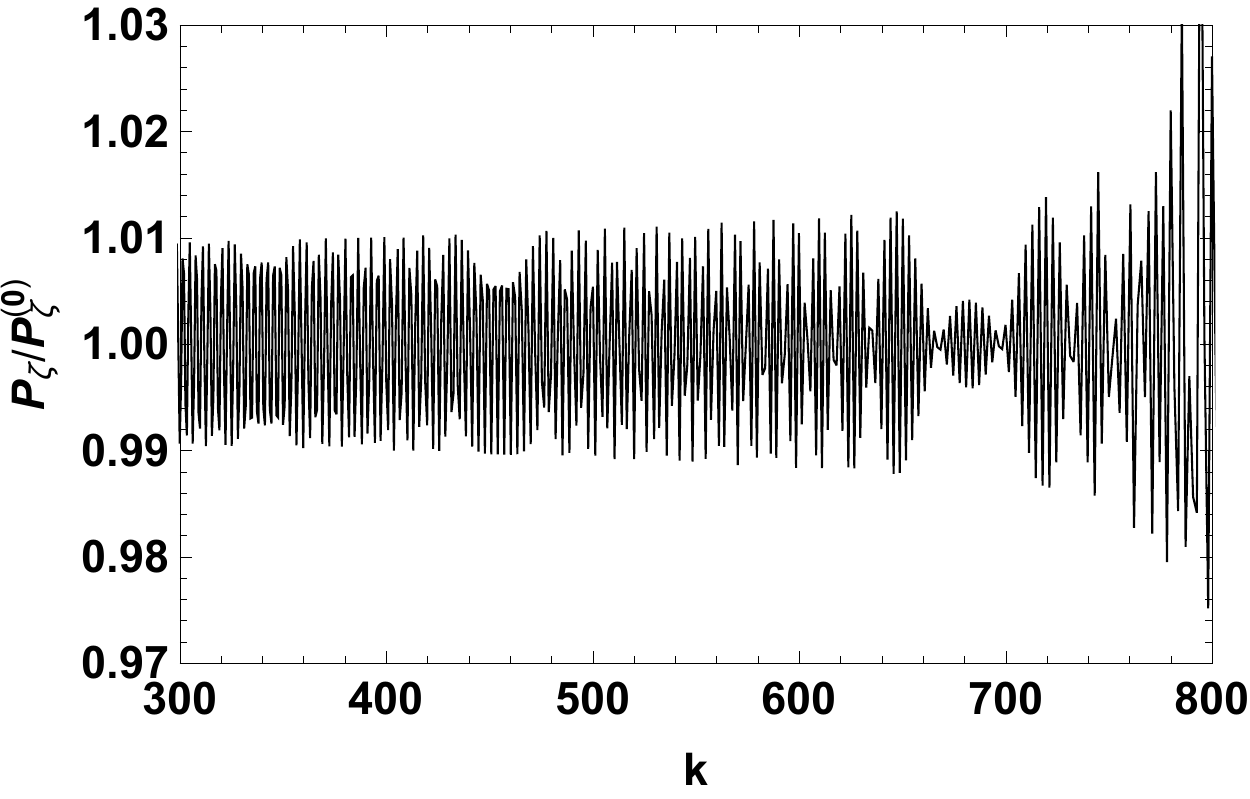}   \includegraphics[width=0.32\textwidth]{model1plot5}   \\     
    \caption{\label{fig:comparison2} The first column corresponds to $\frac{M^2}{f^2} \chi^2 h^2$ model with $h_0 = 800 H$; the last column corresponds to $\frac{M^2}{f} \chi h^2$. The other parameters are the same, $M = 1020H, m_\chi = 10 H, \Lambda = 6000 H$ and $\lambda = 1$. The first row is the Higgs field evolution. The second to last rows correspond to change in the inflaton two-point function in the $k$ range: $(10 - 60), (60 - 100), (150 -300)$, and $(300 - 800)$. }
    \end{center} 
\end{figure}

Now consider that the Higgs obtains its mass through a coupling to the modulus 
\begin{align}
V \supset \frac{M^2}{2f^2} \chi^2 h^2.
\end{align}
As the modulus oscillates, the Higgs is always in the unbroken phase yet with a varying potential. Analogous to case {\it a)} in Sec.~\ref{sec:compare}, the Higgs oscillates around the minimum with period set by $(M |\cos (m_\chi t)|)^{-1}$. The resulting correction to the primordial spectrum is presented in Fig.~\ref{fig:comparison2}. We also include the spectrum for the phase oscillation model for comparison. The spectra are quite different in the full $k$ range. In the symmetric model, the oscillations have a fixed period $1/H$ at low $k$ and appear irregular at large $k$ without a clear k-wavepacket feature. 

\bibliography{ref}
\bibliographystyle{jhep}

\end{document}